\documentclass[aps,prd,groupedaddress,twocolumn,floatfix,letterpaper]{revtex4-2}

\usepackage{anyfontsize}
\usepackage[hidelinks]{hyperref}
\usepackage{graphicx}
\usepackage{amsmath}
\usepackage{bm}
\usepackage{lipsum}
\usepackage{color}
\usepackage{lineno}

\newcommand{\dd}{{\rm d}}
\newcommand{\beq}{\begin{equation}} 
\newcommand{\eeq}{\end{equation}}
\newcommand{\bqa}{\begin{eqnarray}}
\newcommand{\eqa}{\end{eqnarray}}

\begin{document}

\title{Bottomonium suppression in 5.02 and 8.16 TeV \texorpdfstring{$p\,$}{p}-Pb collisions}
\author{Michael Strickland}
\author{Sabin Thapa}
\affiliation{Kent State University, Department of Physics, Kent, Ohio 44242 USA}
\author{Ramona Vogt}
\affiliation{Nuclear and Chemical Sciences Division, Lawrence Livermore National Laboratory, Livermore, California 94551, USA}
\affiliation{Physics and Astronomy Department, University of California at Davis, Davis, California 95616, USA}

\begin{abstract}
We compute the suppression of $\Upsilon(1S)$, $\Upsilon(2S)$, and $\Upsilon(3S)$ states in $p$-Pb collisions relative to $pp$ collisions, including nuclear parton distribution function (nPDF) effects, coherent energy loss, momentum broadening, and final-state interactions in the quark-gluon plasma.  We employ the EPPS21 nPDFs and calculate the uncertainty resulting from variation over the associated error sets.  To compute coherent energy loss and momentum broadening, we follow the approach of Arleo, Peigne, and collaborators.  The 3+1D viscous hydrodynamical background evolution of the quark-gluon plasma is generated by anisotropic hydrodynamics.  The in-medium suppression of bottomonium in the quark-gluon plasma is computed using a next-to-leading-order open quantum system framework formulated within potential nonrelativistic quantum chromodynamics.  We find that inclusion of all these effects provides a reasonable description of experimental data from the ALICE, ATLAS, CMS, and LHCb Collaborations for the suppression of $\Upsilon(1S)$, $\Upsilon(2S)$, and $\Upsilon(3S)$ as a function of both transverse momentum and rapidity.
\end{abstract}


\maketitle

%%%%%%%%%%%%%%%%%%%%%%%%%%%%%%%%%%%%%%%%%%%%%%%%%%%%%%%%%%%%%%%%
\section{Introduction}
%%%%%%%%%%%%%%%%%%%%%%%%%%%%%%%%%%%%%%%%%%%%%%%%%%%%%%%%%%%%%%%%

The theoretical and experimental study of heavy quarkonium suppression in $AA$ and $pA$ collisions can be used to understand the role of both cold and hot nuclear matter effects on strongly bound states.  Because both cold and hot suppression mechanisms can be relevant, it is necessary to have a comprehensive model of both in order to draw quantitative conclusions about the different sources of heavy quarkonium suppression in such collisions.

Almost forty years ago, Matsui and Satz proposed that suppression of heavy quarkonium production in $AA$ collisions relative to their production in $pp$ collisions could be used to assess whether or not a deconfined quark-gluon plasma (QGP) was created in such collisions \cite{Matsui:1986dk}.  Their predictions were based on the expectation that the production of a QGP would reduce the binding energy of heavy quarkonium states through the color Debye screening of the interaction between quarks and antiquarks in the deconfined phase of quantum chromodynamics (QCD)~\cite{Gross:2022hyw}.  In the Matsui-Satz picture, excited states would experience higher suppression than the ground state and this effect would provide a way to gauge whether or not a QGP had been formed and infer the temperature of the produced QGP \cite{Mocsy:2013syh,Andronic:2015wma}.

In recent years, however, there has been a paradigm shift in our understanding of the temperature dependence of the heavy quark-antiquark potential, indicating that color screening may not be the driving force behind the breakup of heavy quarkonium in the QGP.  The widths of the heavy quarkonium states were determined to be large at temperatures above the deconfinement temperature, causing the breakup of these states at temperatures below their naive dissociation temperatures \cite{Laine:2006ns,Beraudo:2007ky,Escobedo:2008sy,Dumitru:2009fy,Brambilla:2008cx,Brambilla:2010vq,Margotta:2011ta,Brambilla:2011sg,Brambilla:2013dpa}.  These large in-medium widths were reflected by the emergence of imaginary-valued contributions to the heavy quark-antiquark potential.   Resummed perturbative and effective field theory calculations of the imaginary part of the heavy-quark potential have been confirmed by first-principles nonperturbative Euclidean lattice QCD and classical real-time lattice QCD calculations~\cite{Rothkopf:2011db,Petreczky:2018xuh,Bala:2019cqu,Laine:2007qy,Lehmann:2020fjt,Boguslavski:2020bxt}.  The most recent lattice QCD calculations indicate that heavy quarkonium bound states possess large widths and that there is very little modification of the real part of the potential \cite{Larsen:2019bwy,Bala:2021fkm}.  This result provides a fundamental challenge to the original Matsui-Satz picture and forces us to consider the QGP-induced breakup of heavy quarkonium states at temperatures lower than their naive dissociation temperatures, for example, at temperatures generated in $pA$ collisions.

Recent theoretical and phenomenological studies have provided a self-consistent quantum mechanical description of heavy quarkonium evolution in the QGP in terms of an open quantum system (OQS) in which the heavy-quark bound state is coupled to a fluctuating thermal medium~\cite{Katz:2015qja,Brambilla:2016wgg,Brambilla:2017zei,Blaizot:2017ypk,Yao:2018nmy,Blaizot:2018oev,Brambilla:2020qwo,Yao:2020xzw,Yao:2020eqy,Brambilla:2021wkt,Yao:2021lus,Blaizot:2021xqa,Miura:2022arv,Brambilla:2022ynh,Scheihing-Hitschfeld:2022xqx,Katz:2022fpb,Brambilla:2023hkw,Nijs:2023dks,Scheihing-Hitschfeld:2023tuz,Nijs:2023bok}.  One of the conceptual issues brought forth by such studies was how to formulate the time evolution of the reduced density matrix of the heavy quark-antiquark pairs in a manner that preserved unitary and positivity.  The solution to this problem came through the formulation of the problem as an OQS in which heavy-quark-antiquark pairs are described in terms of a reduced density matrix that is obtained by integrating over the medium degrees of freedom. For recent reviews, see Refs.~\cite{Rothkopf:2019ipj,Akamatsu:2020ypb,Yao:2021lus}.

Because heavy quarkonium widths can be on the order of a few hundred MeV, QGP-induced heavy quarkonium breakup can occur on timescales that are short compared to the QGP lifetime and, therefore, must be taken into account in phenomenological applications. This understanding has been used in many phenomenological studies of heavy quarkonium suppression in the QGP generated in $AA$ collisions~\cite{Strickland:2011mw,Strickland:2011aa,Krouppa:2015yoa,Islam:2020bnp,Islam:2020gdv,Brambilla:2020qwo,Brambilla:2021wkt,Brambilla:2022ynh,Wen:2022yjx,Alalawi:2022gul,Strickland:2023nfm,Andronic:2024oxz}.

Despite the concerted effort to understand heavy quarkonium suppression in the QGP, key theoretical uncertainties related to the role of cold nuclear matter (CNM) effects, such as the modification of parton distribution functions in nuclear environments (nPDFs) \cite{Vogt:2022glr,Eskola:2021nhw,Kovarik:2015cma,AbdulKhalek:2022fyi}, energy loss and momentum broadening of the produced quarkonia \cite{Arleo:2012rs,Arleo:2013zua,Liou:2014rha,Arleo:2014oha}, and interactions with comoving hadrons 
\cite{Gavin:1988hs,Vogt:1988fj,Gavin:1990gm,Capella:1996va,Ftacnik:1988qv,Armesto:1997sa,Capella:2000zp,Capella:2005cn,Capella:2006mb,Capella:2007jv,Ferreiro:2012rq,Ferreiro:2014bia} remain~\footnote{Nuclear absorption of heavy quarkonium is found to be negligible at LHC energies~\cite{Lourenco:2008sk}.}.  To have a firm quantitative understanding of the role played by both cold and hot nuclear matter effects, it is necessary to systematically analyze heavy quarkonium suppression in both $AA$ and $pA$ collisions.  In the former, CNM effects are expected to be subleading, resulting in a suppression on the order of 20\%-30\% compared to $pp$ collisions ($R_{\rm AA} \sim 70-80\%$), which is insufficient to explain, for example, the large suppression of bottomonium production observed at the Large Hadron Collider (LHC) \cite{CMS:2017ycw,Sirunyan:2018nsz,ALICE:2019pox,Acharya:2020kls,Lee:2021vlb,CMS:2020efs,ATLAS:2022exb,CMS:2023lfu}. 

However, at lower collision energies, it may be necessary to include CNM effects in order to quantitatively understand the experimental observations made at Brookhaven National Laboratory's Relativistic Heavy Ion Collider \cite{Strickland:2023nfm,STAR:2013kwk,PHENIX:2014tbe,STAR:2016pof}. Additionally, in $pA$ collisions at LHC energies, the suppression of $\Upsilon(1S)$ production is understood to come primarily from nPDF, energy loss, and momentum-broadening effects.  This makes it critical to include these effects to understand the transverse momentum and rapidity dependence of the suppression observed in such collisions.  Experimental observations of excited bottomonium states such as the $\Upsilon(2S)$ and $\Upsilon(3S)$ indicate that these states are more suppressed than the ground state. This differential suppression cannot be understood solely in terms of nPDF and energy loss and momentum broadening.  Typically, final-state interactions in the framework of comover models are invoked to explain the greater suppression of heavy quarkonium excited states relative to the ground state~\cite{Capella:1996va,Armesto:1997sa,Capella:2000zp,Capella:2005cn,Capella:2006mb,Capella:2007jv,Ferreiro:2012rq,Ferreiro:2014bia}; however, this is not the only possible final-state interaction that must be taken into account.  One must also consider the possibility that there is a short-lived QGP created in $p$-Pb collisions \cite{Strickland:2018exs,Noronha:2024dtq}, which can cause final-state suppression of heavy quarkonium~\footnote{Traditionally, comover models described comoving hadrons; however, recently such models have opened up the possibility that the comovers could also be deconfined partons.  In this latter case, one may consider this new class of comover models as a phenomenological model of interactions with a deconfined QGP.}.

This possibility has previously been explored in the literature.  In  Ref.~\cite{Du:2018wsj}, the authors considered a combination of EPS09 nPDF modifications and hot QGP effects on $J/\psi$ and $\psi(2S)$ production using a transport model that included regeneration effects.  The dynamical background was calculated using a 2+1D fireball model with a first-order phase transition applied in different rapidity intervals, assuming different initial temperatures in each interval.  In Ref.~\cite{Dinh:2019ajl}, the authors considered the effect of both cold and hot nuclear matter effects on $\Upsilon$ suppression.  They considered nPDF effects and coherent energy loss and used 2+1D ideal hydrodynamics for the background evolution.
In Ref.~\cite{Wen:2022utn}, the authors considered EPS09 nPDFs and hot QGP effects on $J/\psi$ and $\psi(2S)$ production using a complex potential model that included screening and in-medium decays.  They coupled the quantum mechanical evolution to a 2+1D ideal hydrodynamics background with a first-order phase transition.  Similarly to the first model, they applied their 2+1D background in separately tuned rapidity intervals.  In Ref.~\cite{Kim:2022lgu}, the authors considered only hot QGP effects on $\Upsilon(nS)$ production in a wide variety of systems, including $pp$, $p$-Pb, $p$-O, and O-O. The underlying model was based on transport theory without regeneration, with dissociation rates calculated assuming gluodissociation and inelastic parton scattering.  The hydrodynamic background used was provided by event-by-event 2+1D viscous hydrodynamics generated by the SONIC code~\cite{Romatschke:2015gxa}.  Finally, in Ref.~\cite{Chen:2023toz}, the authors combined EPS09 nPDFs and hot QGP effects on $\Upsilon(nS)$ production using transport theory and 2+1D ideal hydrodynamics applied only at central rapidity.  In all the references listed above, the authors found that there could be important final-state effects on heavy quarkonium states arising from propagation through a short-lived QGP.
 
In this paper, we present a comprehensive calculation of cold and hot nuclear matter effects on bottomonium production in minimum-bias (min-bias) $p$-Pb collisions.  We employ a three-stage model that takes into account: nPDF effects on initial bottomonium production using EPPS21 nPDF sets~\cite{Eskola:2021nhw}; coherent energy loss and transverse momentum broadening following the formalism of Refs.~\cite{Arleo:2012rs,Arleo:2013zua,Liou:2014rha,Arleo:2014oha}; and the real-time evolution of bottomonium states in the QGP produced in $p$-Pb collisions using an OQS Lindblad equation solver formulated within the potential nonrelativistic QCD (pNRQCD) effective theory \cite{Brambilla:2023hkw}.  Our work goes beyond previous studies of bottomonium suppression in $p$-Pb collisions by including state-of-the-art calculations of all three effects in combination. In addition, we use a 3+1D dissipative hydrodynamic background with a realistic equation of state when computing QGP-induced suppression.  Based on this, we can systematically compute the magnitude of each effect as a function of transverse momentum and rapidity, individually and in combination, allowing comparisons with experimental data collected in different rapidity intervals by the ALICE, ATLAS, CMS, and LHCb Collaborations in a manner consistent with the transverse momentum and rapidity dependence of soft-hadron production in \mbox{$p$-Pb} collisions~\cite{ALICE:2019qie,ATLAS:2017prf,CMS:2022wfi,LHCb:2018psc}.

We find that the inclusion of these three effects allows for a quite reasonable description of available data given current experimental and theoretical uncertainties. Final-state effects on the $\Upsilon(1S)$ represent a small correction relative to the effects of nPDFs, coherent energy loss, and transverse momentum broadening. However, final-state effects on $\Upsilon(2S)$ and $\Upsilon(3S)$ production are necessary to obtain agreement with the available data.  This necessity provides further evidence for the production of a hot, but short-lived, QGP being generated in 
\mbox{$p$-Pb} collisions.  This conclusion supports previous studies of quarkonium suppression  in \mbox{$p$-Pb} collisions \cite{Du:2018wsj,Dinh:2019ajl,Wen:2022utn,Kim:2022lgu,Chen:2023toz} that found that inclusion of final-state interactions in the QGP resulted in increased suppression of excited heavy quarkonium states relative to their ground states.

The structure of our paper is as follows. In Sec.~\ref{sec:npdf}, we provide details concerning nPDF effects on bottomonium production.  In Sec.~\ref{sec:eloss}, we discuss the effects of energy loss and transverse momentum broadening.  In Sec.~\ref{sec:qgp}, we provide details of the OQS+pNRQCD framework and the 3+1D hydrodynamical evolution used as the background for computing bottomonium suppression in the QGP.  In Sec.~\ref{sec:combo}, we discuss how initial- and final-state effects are combined.  In Sec.~\ref{sec:feeddown}, we describe how late-time feed down of bottomonium excited states is included.  In Sec.~\ref{sec:results}, we present our final results compared to experimental data from the ALICE, ATLAS, CMS, and LHCb Collaborations.  In Sec.~\ref{sec:conclusions}, we present our conclusions and an outlook for the future.  Finally, in the Appendix, we provide details concerning the dependence of all effects on $\sqrt{s_{NN}}$.

%%%%%%%%%%%%%%%%%%%%%%%%%%%%%%%%%%%%%%%%%%%%%%%%%%%%%%%%%%%%%%%%
\section{Nuclear parton distribution function effects}
\label{sec:npdf}
%%%%%%%%%%%%%%%%%%%%%%%%%%%%%%%%%%%%%%%%%%%%%%%%%%%%%%%%%%%%%%%%

We begin by presenting the calculation of $\Upsilon$ production within the color evaporation model (CEM) \cite{Gavai:1994in}.  This model, together with the improved color evaporation model \cite{Ma:2016exq}, can describe the $\Upsilon$ rapidity and transverse momentum distributions, including at rather low $p_T$.

The CEM assumes that some fraction, $F_C$, of the $b \overline b$ pairs
produced with a pair mass below the $B \overline B$ pair mass threshold
will go on mass shell as an $\Upsilon$ state,
\bqa
\sigma_{\rm CEM}(pp) & = & F_C \sum_{i,j} 
\int_{4m^2}^{4m_H^2} d\hat{s}
\int dx_1 \, dx_2~ \label{sigCEM} \\ 
&& \hspace{-9mm} \times F_i^p(x_1,\mu_F^2,k_{T_1})~ F_j^p(x_2,\mu_F^2,k_{T_2})~ 
\hat\sigma_{ij}(\hat{s},\mu_F^2, \mu_R^2) \, . \nonumber
%\label{sigCEM}
\eqa
where $ij = g+g, q + \overline q$, or $q(\overline q)+ g$ and
$\hat\sigma_{ij}(\hat {s},\mu_F^2, \mu_R^2)$
is the partonic cross section for initial state $ij$ with the
$q(\overline q)+g$ process appearing at next-to-leading order (NLO) in $\alpha_s$.  
The cross section and parton distribution functions are calculated at factorization 
scale $\mu_F$ and renormalization scale $\mu_R$. The NLO heavy-flavor
cross section is obtained using the HVQMNR code \cite{Mangano:1991jk}.
The values for the bottom quark
mass, $M$, and the scales $\mu_F$ and $\mu_R$,
are determined from a fit to the total $b \overline b$
cross section at NLO:
$(m,\mu_F/m_T, \mu_R/m_T) = (4.65 \pm 0.09 \, {\rm GeV}, 1.40^{+0.77}_{-0.59}, 1.10^{+0.22}_{-0.20})$.
The scales are defined relative to the transverse mass of the pair,
\mbox{$\mu_{F,R} \propto m_T = \sqrt{M^2 + p_T^2}$}, where 
the $p_T$ is the $b \overline b$ pair $p_T$, 
\mbox{$p_T^2 = 0.5(p_{T,Q}^2 + p_{T,{\overline Q}}^2)$}.  
The normalization factor
$F_C$ is obtained by fitting the energy dependence of the summed $\Upsilon$($n$S)
cross sections at $y=0$, multiplied by the branching ratio to muon pairs.
The normalizations of individual states are determined based on the individual
branching ratios and feed down.

The parton densities in Eq.~\eqref{sigCEM} include intrinsic $k_T$, required
to keep the pair cross section
finite as $p_T \rightarrow 0$.
They are assumed to factorize into the normal collinear parton densities and a $k_T$-dependent function,
\bqa
F^p(x,\mu_F^2,k_T) = f^p(x,\mu_F^2)G_p(k_T) \, . \label{PDFfact}
\eqa
The CT10 proton parton distribution functions (PDFs)
\cite{Lai:2010vv} are employed in the calculations of $f^p(x,\mu_F^2)$.

At leading order (LO) in the CEM, the $Q \overline Q$ pair $p_T$ is zero.  Thus, $k_T$ broadening is required to keep the $p_T$ distribution finite as $p_T \rightarrow 0$. Broadening has typically been modeled by intrinsic transverse momentum, $k_T$, added to the parton densities and playing the role of low transverse momentum QCD resummation \cite{Lo:1979he}.  

In the HVQMNR code, an intrinsic $k_T$ is added to each final state bottom quark, 
rather than to the initial state, as in the case of
Drell-Yan production \cite{Lo:1979he}. 
In the initial state, the intrinsic $k_T$ function multiplies the parton
distribution functions for both hadrons, 
assuming the $x$ and $k_T$ dependencies factorize,
as in Eq.~\eqref{PDFfact}.  
At leading order, there is no difference between an initial-state (on the partons) or 
final-state (on the produced bottom quarks) $k_T$ kick.  
However, at NLO, when there is a
light parton in the final state, the correspondence can be inexact.  
The difference between the two implementations is small if
$\langle k_T^2 \rangle \leq 2-3$ GeV$^2$ \cite{Mangano:1992kq}.

A Gaussian distribution is employed for
$G_p(k_T)$ in Eq.~\eqref{PDFfact}~\cite{Mangano:1992kq}, 
\bqa
G_p(k_T) = \frac{1}{\pi \langle k_T^2 \rangle_p} \exp(-k_T^2/\langle k_T^2
\rangle_p) \, .
\label{intkt}
\eqa
The rapidity distributions are independent of the intrinsic $k_T$.

The broadening is applied by boosting the transverse momentum
of the $b \overline b$ pair
(plus light parton at NLO)
to its rest frame
from the longitudinal center-of-mass frame.
The transverse momenta of the incident partons, $\vec k_{T1}$ and
$\vec k_{T2}$, or, in this case, the final state $b$ and $\overline b$ quarks,
are redistributed isotropically with unit modulus, according to
Eq.~\eqref{intkt}, preserving momentum conservation.
Once boosted back to the initial frame, the transverse momentum of
the $b \overline b$ pair changes from $\vec p_T$ to
$\vec p_T + \vec k_{T 1} + \vec k_{T 2}$~\cite{Frixione:1994nb}.  

The broadening effect 
decreases as $\sqrt{s_{NN}}$ increases because the perturbatively calculated average
$p_T$ of the pair $b \overline b$ also increases
with $\sqrt{s_{NN}}$.  The value of $\langle k_T^2 \rangle_p$ is assumed to
increase with $\sqrt{s_{NN}}$ so that the effect is non-negligible for low $p_T$ heavy-flavor production at higher energies.
The energy dependence of $\langle k_T^2 \rangle_p$ in Ref.~\cite{Nelson:2012bc} is
\bqa
  \langle k_T^2 \rangle_p = \left[ 1 + \frac{1}{n} \ln
    \left(\frac{\sqrt{s_{NN}} \, ({\rm GeV})}{20 \,
    {\rm GeV}} \right) \right] \, \, {\rm GeV}^2 \, , 
\label{eq:avekt}
\eqa
with $n = 3$ for $\Upsilon$ production.  Thus,
$\langle k_T^2 \rangle_p$ increases
rather slowly with energy.  At the LHC $pA$ energies, 
\mbox{$\langle k_T^2 \rangle_p = 2.84$~GeV$^2$} and 3~GeV$^2$ for $\sqrt{s_{NN}} = 5.02$ and 8.16~TeV,
respectively.  
The values of $\langle k_T^2 \rangle_p$ are approximately within the range
of applicability proposed in Ref.~\cite{Mangano:1992kq}.

The nuclear modifications of the parton densities are included in the calculation of $\Upsilon$ production in $pA$ collisions as
\bqa
\sigma_{\rm CEM}(pA) & = & F_C \sum_{i,j} 
\int_{4m^2}^{4m_H^2} d\hat{s}
\int dx_1 \, dx_2~ \label{sigCEM_pA} \\ 
&& \hspace{-8mm} \times F_i^p(x_1,\mu_F^2,k_T)~ F_j^A(x_2,\mu_F^2,k_T)~ 
\hat\sigma_{ij}(\hat{s},\mu_F^2, \mu_R^2) \, , \nonumber
%\label{sigCEM_pA}
\eqa
where
\bqa
F_j^A(x_2,\mu_F^2,k_T) & = & R_j(x_2,\mu_F^2,A) F_j^p(x_2,\mu_F^2,k_T) \, .
\eqa
The $k_T$ broadening in the nuclear target is implemented in the final state according to the method described in the next section.  No nuclear absorption is assumed for $\Upsilon$ suppression at the LHC.

The factor $R_j(x_2,\mu_F^2,A)$ represents the nuclear modification of the parton
distributions.  A number of global analyses have been made to
describe the modification as a function of $x$ and the factorization scale $\mu_F$,
assuming collinear factorization and starting from a minimum scale, $\mu_F^0$.
The nPDF effects generated in this scheme are
generally implemented by a parametrization as
a function of $x$, $\mu_F$, and $A$.  

The NLO EPPS21 \cite{Eskola:2021nhw} nPDF parametrization is used in our calculations.
EPPS21 has 24 fit parameters, for
49 total sets: one central set and 48 error sets.  The error sets are
determined by individually varying each parameter within 1 standard deviation
of its best-fit value.
The uncertainties on $R_j(x_2,\mu_F^2,A)$ are calculated by summing the
excursions of each of the error sets from the central value in quadrature. 

The nPDF uncertainties on the $\Upsilon$ distributions are
obtained by calculating the perturbative cross sections at the central values assumed for
the bottom mass and the factorization and renormalization scales employing the central EPPS21 set
as well as the 48 error sets and summing the differences in quadrature.  The deviations of the resulting uncertainty bands from the central cross section are on the order of 20\%.

\begin{figure}
  \begin{center}
    \includegraphics[trim = 100 250 100 200, clip, width=\linewidth]{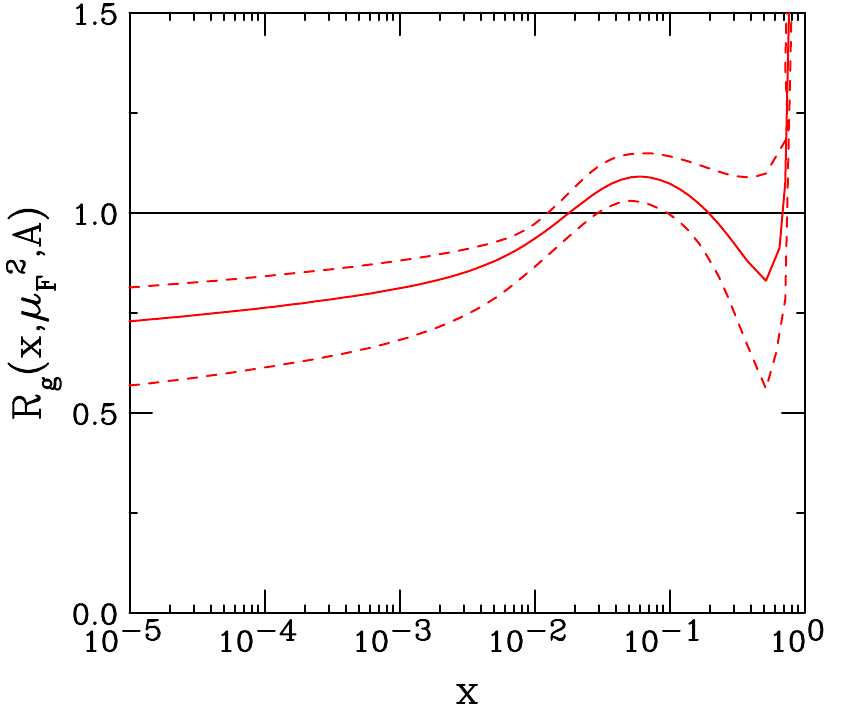}    
  \end{center}
  \vspace{-5mm}
  \caption[]{The EPPS21 ratios, with uncertainties,
    are shown at the scale of the $\Upsilon(1S)$ mass for gluons as a function of
    momentum fraction $x$. The central set is denoted by the solid curve,
    while the dashed curves give the upper and lower limits of the uncertainty
    bands.  
  }
\label{fig:shad_ratios}
\end{figure}

The EPPS21 ratios for gluons are
shown at the $\Upsilon$ mass scale in Fig.~\ref{fig:shad_ratios}.
The central sets, along with the uncertainty
bands, are shown for Pb ($A=208$) targets.  The antishadowing peak around
$x\sim 0.05$ is at backward rapidity in $p$-Pb collisions, while low-$x$ shadowing is manifested at forward rapidity.

%%%%%%%%%%%%%%%%%%%%%%%%%%%%%%%%%%%%%%%%%%%%%%%%%%%%%%%%%%%%%%%%
\section{Energy loss and momentum broadening}
\label{sec:eloss}
%%%%%%%%%%%%%%%%%%%%%%%%%%%%%%%%%%%%%%%%%%%%%%%%%%%%%%%%%%%%%%%%

Quarkonium states can also undergo an elastic scattering, thereby exchanging a gluon with cold nuclear matter or the nuclear target, and this scattering can induce radiation. The medium-induced radiation spectrum caused by the gluon radiation is coherent. This coherent energy loss is different from the gluon radiation resummed in leading-twist parton distribution and fragmentation functions and should also be considered~\cite{Arleo:2012rs,Liou:2014rha}. Taking into account coherent parton energy loss $\varepsilon$ in cold nuclear matter and transverse-momentum broadening $\delta p_T$, the quarkonium double differential cross section in $pA$ collisions can be written in terms of the $pp$ quarkonium production cross section~\cite{Arleo:2013zua},
\bqa
\frac{1}{A}\frac{\dd\sigma_{pA}^{\Upsilon}}{\dd y \, \dd^2 {\vec p}_T}(y,{\vec p}_T) &=& \int_\varphi \int_0^{\varepsilon_{\rm max}(y)} \dd \varepsilon \,{\cal P}(\varepsilon, E(y), \ell^2) \nonumber \\
&& \hspace{-2.5cm} \times \left[ \frac{E(y)}{E(y)+\varepsilon} \right] \frac{\dd\sigma_{pp}^{\Upsilon}}{\dd y \, \dd^2 {\vec p}_T} \left( y(E+\varepsilon), | \vec{p}_T - \delta \vec{p}_T | \right) , \;
\label{eq:eloss1}
\eqa
where $y$ is the momentum rapidity, ${\vec p}_T$ is the transverse momentum, and \mbox{$\int_\varphi = \int_{0}^{2\pi} \dd \varphi/(2\pi)$}. In Eq.~\eqref{eq:eloss1}, $E$ is the quarkonium energy and $\varepsilon$ is the energy loss experienced by the state as it propagates through the nucleus in the nuclear rest frame. The quenching weight, $\cal P$, is the probability density for the energy loss as a function of the radiated energy and the energy of the radiating particle. The upper limit on the integral over $\varepsilon$ is given by $\varepsilon_{\rm max} = {\rm min}(E_p-E,E)$ where $E_p \simeq s/(2m_p)$ is the proton energy in the nuclear rest frame~\cite{Arleo:2012rs}, $s$ is the Mandelstam variable, and $m_p$ is the mass of the proton.

The magnitude of the transverse momentum broadening in $pA$ collisions can be expressed as $\delta p_T = \sqrt{\ell^2_A - \ell^2_p}$,  where $\ell_A$ and $\ell_p$ are the transverse momentum broadening due to the traversal of the quarkonium state through a nucleus or a proton, respectively.  The nuclear momentum broadening is given in terms of the transport coefficient $\hat{q}$ times the effective path length $L_A$ through the target nucleus $A$, {\it i.e.},  $\ell^2_A = \hat{q} L_A $, with $\hat{q}$ depending on the kinematics, as specified below.

In Eq.~\eqref{eq:eloss1}, the integral over the azimuthal angle is the integral over azimuthal angle$\varphi$ of $\delta \vec{p}_T$, which is assumed to be uniformly distributed in the transverse plane.  In this case, the transverse momentum $\vec{p}_T$ can be taken along the $x$ axis by choosing a coordinate system such that $\vec{p}_T = (p_T,0)$ and $\delta \vec{p}_T = \delta p_T (\cos\varphi,\sin\varphi)$, giving $| \vec{p}_T - \delta \vec{p}_T | = \sqrt{(p_T - \cos\varphi \, \delta p_T)^2 + (\sin\varphi \, \delta p_T)^2}$. 

We note that, by making a change of variables, one can rewrite Eq.~\eqref{eq:eloss1} in terms of the rapidity shift $\delta y$ defined as
\beq
E(y) + \varepsilon \equiv E(y + \delta y) = E(y) e^{\delta y} \, ,
\eeq
which gives
\bqa
\frac{1}{A}\frac{\dd\sigma_{\mathrm{pA}}^{\Upsilon}}{\dd y \, \dd^2 {\vec p}_T}(y,{\vec p}_T) &=& \int_\varphi \int_0^{\delta y_{\rm max}(y)} \dd \delta y \,\hat{\cal P}(e^{\delta y} - 1, \ell^2) \nonumber \\
&& \hspace{-2mm} 
\times \frac{\dd\sigma_{\mathrm{pp}}^{\Upsilon}}{\dd y \, \dd^2 {\vec p}_T} \left( y + \delta y, | \vec{p}_T - \delta \vec{p}_T | \right) ,
\label{eq:elossfinal}
\eqa
with $\delta y_{\rm max} = \text{min}(\ln 2, y_{\rm max}-y)$, where \mbox{$y_{\rm max} = \ln(\sqrt{s_{NN}}/m_{T,\Upsilon})$} is the maximum bottomonium rapidity in the proton-nucleon center-of-momentum frame and \mbox{$m_{T,\Upsilon} = \sqrt{p_T^2 + M_\Upsilon^2}$} is the transverse mass of the bottomonium state.

%%%%%%%%%%%%%%%%%%%%%%%%%%%%%%%%%%%%%%%%%%%%%%%%%%%%%%%%%%%%%%%%
\subsection{The coherent energy loss quenching weight}
%%%%%%%%%%%%%%%%%%%%%%%%%%%%%%%%%%%%%%%%%%%%%%%%%%%%%%%%%%%%%%%%

The quenching weight ${\cal P}$ can be obtained from the radiation spectrum ${\dd I}/{\dd \omega}$.  The fully coherent energy loss result is~\cite{Arleo:2012rs}
\bqa
\label{relation-P-spec}
{\cal P}(\varepsilon, E, \ell_A^2) &=& \frac{\dd I}{\dd\varepsilon} \, \exp \left\{ - \int_{\varepsilon}^{\infty} \dd\omega  \frac{\dd{I}}{\dd\omega} \right\} \nonumber \\
&=& \frac{\partial}{\partial \varepsilon} \, \exp \left\{ - \int_{\varepsilon}^{\infty} \dd\omega  \frac{\dd{I}}{\dd\omega} \right\} ,
\eqa
with
\bqa
\label{eq:spectrum-def}
\frac{\dd I}{\dd \omega} &=& \frac{N_c \alpha_s}{\pi \omega} \left\{ \ln\!{\left(1+\frac{\ell_A^2 E^2}{m_{T,\Upsilon}^2 \omega^2}\right)} - \ln\!{\left(1+\frac{\Lambda_{p}^2 E^2}{m_{T,\Upsilon}^2 \omega^2}\right)} \right\} \nonumber \\
&& \hspace{4cm} \times \Theta(\ell_A^2 - \Lambda_{p}^2) \, .
\eqa
The quantity \mbox{$\Lambda_{p}^2 = {\rm max}(\Lambda_{\mathrm{QCD}}^2,\ell_{p}^2)$} is the quarkonium momentum broadening experienced when traversing a proton, with QCD scale parameter $\Lambda_{\rm QCD}$.  Above, $N_c$ is the number of colors and $\alpha_s$ is the QCD coupling constant evaluated at the average scale of the momentum transfer.

From the above expression, one finds that $E\,{\cal P}(\varepsilon,E,\ell_A^2)$ scales as a function of $x \equiv \varepsilon/E$ and $\ell_A^2$.  This allows one to introduce a scaled function $\hat{\cal P}$ defined as
$\hat{\cal P}(x,\ell_A^2) \equiv E\,{\cal P}(\varepsilon,E,\ell_A^2)$, which can be expressed in terms of the dilogarithm function ${\rm Li}_2(x)$ as~\cite{Arleo:2013zua}
\bqa
\hat{\cal P}(x,\ell_A^2) &=&  \frac{\partial}{\partial x} \, \exp \left\{\frac{N_c \alpha_s}{2 \pi} \left[ {\rm Li}_2\!\left(\frac{-\ell_A^2}{x^2 m_{T,\Upsilon}^2} \right) \right. \right.
\nonumber \\
&& \hspace{1.2cm} \left. \left.\hspace{1.5cm} - {\rm Li}_2\!\left(\frac{-\Lambda_{p}^2}{x^2 m_{T,\Upsilon}^2} \right) \right] \right\} . \;\;\;\;\;
\label{eq:quenching-dilog}
\eqa
%
%%%%%%%%%%%%%%%%%%%%%%%%%%%%%%%%%%%%%%%%%%%%%%%%%%%%%%%%%%%%%%%%
%\subsection{Fixing the parameters entering into the quenching weight}
%%%%%%%%%%%%%%%%%%%%%%%%%%%%%%%%%%%%%%%%%%%%%%%%%%%%%%%%%%%%%%%%

We take $N_c = 3$ and use the four-loop running for the strong coupling constant $\alpha_s$, with the scale given by $\ell_A$ and $\Lambda_{\rm QCD} = 0.308$ GeV.  This value of $\Lambda_{\rm QCD}$ gives $\alpha_s(1.5\;{\rm GeV}) = 0.326$, which is the value extracted from lattice QCD calculations of the static energy~\cite{Bazavov:2012ka}.  We take \mbox{$M = 9.95$ GeV}, which is the average of the $\Upsilon(1S)$, $\Upsilon(2S)$, and $\Upsilon(3S)$ masses~\footnote{We have explicitly verified that using distinct masses for the different $\Upsilon$ states results in very small variation in the coherent energy loss (line thickness in plots).}.

Based on fits to HERA data \cite{Golec-Biernat:1998zce}, we assume
\bqa
&& \hat{q} = \hat{q}_0 \left( \frac{10^{-2}}{x_A} \right)^{0.3} \, , \nonumber \\
&& x_A = {\rm min}(x_0,x_2) \, , \nonumber \\
&& x_{0} = 1/(2 m_p L_A) \, , \nonumber \\
&& x_2 = \frac{m_T}{\sqrt{s_{NN}}} e^{-y}  \, .
\label{eq:eqs}
\eqa
The parameter \mbox{$\hat{q}_0 = \hat{q}(x=10^{-2})$} is fixed to experimental data.  We follow Arleo and Peigne \cite{Arleo:2012rs} and use
\beq
\hat{q}_0 = 0.075 \; {\rm GeV}^2/{\rm fm} \, .
\eeq

To compute the effect of momentum broadening, an estimate of the effective path length traversed by the parton through the nucleus is required. We use~\cite{Arleo:2012rs}
\beq
L_{A} = L_{p} + \frac{A-1}{A^2 \rho_0} \int \dd^2 \vec b \, T_A^2(\vec b) \, ,
\label{eq:leff}
\eeq
where $T_A$ is the nuclear thickness function for $A = 208$ and $\vec b$ is impact parameter.  To compute $T_A$, we assume a Woods-Saxon distribution for the nucleon density with $\rho_0 = 0.17$ fm$^{-3}$, $d = 0.54$ fm, and \mbox{$R_{\rm Pb} = 6.49$ fm}.  The effective path length for the proton is taken to be \mbox{$L_{p} = 1.5 \; {\rm fm}$}~\cite{Arleo:2012rs}.  The effective path length for Pb is calculated using Eq.~\eqref{eq:leff}, giving $L_{\rm Pb} = 10.41 \; {\rm fm}$.

%%%%%%%%%%%%%%%%%%%%%%%%%%%%%%%%%%%%%%%%%%%%%%%%%%%%%%%%%%%%%%%%
\subsection{Parametrization of the \texorpdfstring{$pp$}{pp} cross section}
%%%%%%%%%%%%%%%%%%%%%%%%%%%%%%%%%%%%%%%%%%%%%%%%%%%%%%%%%%%%%%%%

The double differential cross section of the prompt quarkonia entering into Eqs.~\eqref{eq:eloss1} and \eqref{eq:elossfinal} can be parametrized as~\cite{Arleo:2013zua}
\beq
\frac{\dd \sigma^{\Upsilon}_{pp}}{\dd y \, d^2 {\vec p}_T} = \mathcal{N} \left( \frac{p_0^2}{p_0^2 + p_T^2} \right)^\upsilon \left( 1- \frac{2m_T}{\sqrt{s_{NN}}} \cosh y \right)^\beta .
\label{eq:ppdist}
\eeq
Here, the normalization $\mathcal{N}$ is irrelevant since we consider only cross section ratios. The other constants are \mbox{$p_0 = 6.6$ GeV}, $\upsilon = 2.8$, and $\beta = 13.8$, which are obtained from a global fit of $\Upsilon$ production data \cite{Arleo:2013zua}. We have explicitly verified that this fit is in good agreement with the CEM $pp$ cross section introduced in Sec.~\ref{sec:npdf}.  Here, we use this analytic form because it is more efficient in numerical calculations.  In addition, we will use this form when sampling bottomonium transverse momenta and rapidities for propagation through the QGP.

%%%%%%%%%%%%%%%%%%%%%%%%%%%%%%%%%%%%%%%%%%%%%%%%%%%%%%%%%%%%%%%%
\section{Quarkonium suppression in the quark-gluon plasma}
\label{sec:qgp}
%%%%%%%%%%%%%%%%%%%%%%%%%%%%%%%%%%%%%%%%%%%%%%%%%%%%%%%%%%%%%%%%

To understand how heavy quarkonium pairs evolve in the QGP, we use numerical solutions to a Lindblad equation obtained within the pNRQCD effective field theory. The Lindblad equation describes changes in the reduced density matrix of heavy quarkonium in an OQS approach. The results presented here are based on NLO evolution equations, which were obtained by expanding the quantum master equation obeyed by the reduced density matrix to NLO in the binding energy over the temperature~\cite{Brambilla:2022ynh}.

%%%%%%%%%%%%%%%%%%%%%%%%%%%%%%%%%%%%%%%%%%%%%%%%%%%%%%%%%%%%%%%%
\subsection{NLO pNRQCD+OQS}
%%%%%%%%%%%%%%%%%%%%%%%%%%%%%%%%%%%%%%%%%%%%%%%%%%%%%%%%%%%%%%%%

In order to model the final-state interactions experienced by produced bottomonium states, we use a nonequilibrium master equation that can be derived using pNRQCD and OQS methods. We assume the following hierarchy of scales $M \gg 1/a_{0} \gg \pi T \sim m_{D} \gg E_{\rm b}$, where $M$ is the heavy quark mass, $a_{0}$ is the Bohr radius of the quarkonium state, $T$ is the temperature of the medium, $m_{D} \sim gT$ is the Debye mass, and $E_{\rm b}$ is the binding energy of the quarkonium state \cite{Brambilla:2016wgg,Brambilla:2017zei}. 
At NLO in the binding energy over temperature ($E_b/T$), the resulting Lindblad equation can be written as \cite{Brambilla:2022ynh,Strickland:2023nfm,Brambilla:2023hkw}
\bqa
\frac{d\rho(t)}{dt} &=& -i \left[ H, \rho(t) \right] \nonumber \\ && + \sum_{n=0}^1
\left( C_{i}^{n} \rho(t) C^{n\dagger}_{i} - \frac{1}{2} \left\{ C^{n \dagger}_{i} C_{i}^{n}, \rho(t) \right\} \right) , \hspace{4mm}
\label{eq:Lindblad}
\eqa
with the reduced density matrix and Hamiltonian given by
\beq
    \rho(t) = \begin{pmatrix} \rho_{s}(t) & 0 \\ 0 & \rho_{o}(t) \end{pmatrix} 
\eeq
and
\beq
	H = \begin{pmatrix} h_{s} + \text{Im}(\Sigma_{s}) & 0 \\ 0 & h_{o} + \text{Im}(\Sigma_{o}) \end{pmatrix} .
\eeq
The subscripts $s$ and $o$ above refer to singlet and octet states respectively.  The NLO singlet and octet self-energies are given by
\bqa
\label{eq:self_energies_im}
	\text{Im}\left( \Sigma_{s} \right) &=& \frac{r^{2}}{2} \gamma +\frac{\kappa}{4MT} \{r_{i}, p_{i}\} \, , \\
	\text{Im}\left( \Sigma_{o} \right) &=& \frac{N_{c}^{2}-2}{2(N_{c}^{2}-1)} \left( \frac{r^{2}}{2} \gamma +\frac{\kappa}{4MT} \{r_{i}, p_{i}\} \right) .
\eqa
where, in both expressions above, the first terms are LO in $E_b/T$ and the second terms are NLO in $E_b/T$.
The resulting collapse operators, which encode singlet-octet, octet-singlet, and octet-octet transitions, are
\begin{align}
	&\begin{aligned}\label{eq:c0}
	    C_{i}^{0} =& \sqrt{\frac{\kappa}{N_{c}^{2}-1}} \begin{pmatrix} 0 & 1 \\ 0 & 0 \end{pmatrix} \left(r_{i} + \frac{i p_{i}}{2MT} +\frac{\Delta V_{os}}{4T}r_{i} \right) \\ 
	&+ \sqrt{\kappa} \begin{pmatrix} 0 & 0 \\ 1 & 0 \end{pmatrix} \left(r_{i} + \frac{i p_{i}}{2MT} +\frac{\Delta V_{so}}{4T}r_{i} \right) ,
	\end{aligned}\\
	&C_{i}^{1} = \sqrt{\frac{\kappa(N_{c}^{2}-4)}{2(N_{c}^{2}-1)}} \begin{pmatrix} 0 & 0 \\ 0 & 1 \end{pmatrix} \left(r_{i} + \frac{i p_{i}}{2MT} \right) , \label{eq:c1}
\end{align}
where $\Delta V_{os} = V_o - V_s$ is the difference between the octet and singlet potentials, with $V_s = -C_F \alpha_s/r$ and \mbox{$V_o = C_F \alpha_s/{(8r)}$}.  Here, the bottom quark mass is taken to be the $1S$-mass, $M  = $ 4.73 GeV, and the strong coupling constant \mbox{$\alpha_s = g^2/{(4\pi)}$} is evaluated using one-loop running at the scale of the inverse Bohr radius, giving $\alpha_s = 0.468$, following Ref.~\cite{Brambilla:2022ynh}.

The transport coefficients $\kappa$ and $\gamma$ are given by the chromoelectric correlators 
\bqa
\kappa &=& \frac{g^{2}}{18} \int_{0}^{\infty} dt \left\langle \left\{ \tilde{E}^{a,i}(t,\vec{0}), \tilde{E}^{a,i}(0,\vec{0}) \right\} \right\rangle , \\
\gamma &=& -i \frac{g^{2}}{18} \int_{0}^{\infty} dt \left\langle \left[ \tilde{E}^{a,i}(t,\vec{0}), \tilde{E}^{a,i}(0,\vec{0}) \right] \right\rangle  ,
\eqa
where $\tilde{E}$ is a chromoelectric field sandwiched between two links in the fundamental representation, {\it i.e.},
\mbox{$\tilde{E}^{a, i}(t, \vec{0}) = \Omega(t)^\dagger E^{a, i}(t, \vec{0}) \Omega(t)$}, with
\beq
    \Omega(t) = \text{exp}\left[  -ig \int_{-\infty}^{t} \text{d}t' A_{0}(t', \vec{0}) \right] .
\eeq

The Lindblad equation expressed formally in Eq.~(\ref{eq:Lindblad}), with the collapse operators given in Eqs.~\eqref{eq:c0} and \eqref{eq:c1} describes the evolution of the heavy quarkonium reduced density matrix at NLO in $E_{\rm b}/T$~\cite{Brambilla:2022ynh}.  We refer the reader to Ref.~\cite{Brambilla:2022ynh} for details concerning the derivation of these equations.

The transport coefficients $\hat\kappa = \kappa/T^3$ and $\hat\gamma = \gamma/T^3$ can be fixed from direct and indirect lattice calculations.  In Ref.~\cite{Brambilla:2023hkw} it was found that the values of $\hat\kappa=4$ and $\hat\gamma=0$ resulted in a good description of ground- and excited-state bottomonium suppression as a function of the number of participants and transverse momentum in \mbox{$\sqrt{s_{NN}} = $ 5.02 TeV} Pb-Pb collisions.  Later, for \mbox{$\sqrt{s_{NN}} = $ 200 GeV} Au-Au collisions, larger values of $\hat\kappa \sim 5$ were found to be preferred due to the lower temperatures generated at Relativistic Heavy Ion Collider (RHIC) energies~\cite{Strickland:2023nfm}.  This change in $\hat\kappa$ is consistent with observations that $\hat\kappa$ increases as the temperature approaches the QGP phase transition temperature from above~\cite{Altenkort:2023oms}.

Here, we allow $\hat\kappa$ to take even larger values because the average temperature generated in $\sqrt{s_{NN}} = $ 5.02 TeV and 8.16 TeV $p$-Pb collisions is lower than that generated in $\sqrt{s_{NN}} = $ 200 GeV Au-Au collisions.  Again, this is consistent with lattice calculations of the heavy-quark momentum diffusion coefficient in the fundamental representation, which increases in magnitude as the temperature approaches the QGP transition temperature from above~\cite{Altenkort:2023oms}.  The code used to solve the Lindblad equation, QTraj, makes use of the quantum trajectories algorithm and is publicly available \cite{qtraj-download}.  Detailed documentation and code benchmarks are available in Ref.~\cite{Omar:2021kra}.

%%%%%%%%%%%%%%%%%%%%%%%%%%%%%%%%%%%%%%%%%%%%%%%%%%%%%%%%%%%%%%%%
\subsection{3+1D anisotropic hydrodynamics background}
%%%%%%%%%%%%%%%%%%%%%%%%%%%%%%%%%%%%%%%%%%%%%%%%%%%%%%%%%%%%%%%%

To model heavy-ion collisions, one must obtain evolution equations for the resulting 3+1D configurations and include the nonconformality of the QGP consistent with a realistic lattice-based equation of state.  Here, we consider min-bias $p$-Pb collisions, where the lifetime of the QGP is rather short, on the order of 3 to 4 fm/$c$.  Due to the short lifetime of the QGP generated in such events, there can be significant deviations from isotropic equilibrium, and therefore one must take care with both the nonequilibrium evolution and freeze-out \cite{Nopoush:2015yga,Alqahtani:2017mhy}. 

We, therefore, use a far-from-equilibrium formulation of hydrodynamics called anisotropic hydrodynamics~\cite{Martinez:2010sc,Florkowski:2010cf,Alqahtani:2017mhy,Alalawi:2021jwn}, specifically quasiparticle anisotropic hydrodynamics (aHydroQP), in which one assumes that the nonequilibrium QGP consists of massive relativistic quasiparticles with temperature-dependent masses $m(T)$ to take into account the nonconformal nature of the QGP~\cite{Alqahtani:2015qja}. The system is assumed to obey a relativistic Boltzmann equation with $m(T)$ determined from lattice QCD calculations of QCD thermodynamics. Since the masses are temperature dependent, the Boltzmann equation contains an additional force term on the left-hand side related to temperature gradients
\beq
p^\mu \partial_\mu f + \frac{1}{2}\partial_i m^2 \partial_{(p)}^{i} f =  C[f] \, ,
\label{eq:be}
\eeq
where
\beq
C[f] = - \frac{p \cdot u}{\tau_{\rm eq}(T)} [ f - f_\text{eq}(T) ] \, ,
\eeq
is the collisional kernel in the relaxation time approximation. Above, $u^\mu$ is the four-velocity associated with the fluid local rest frame (LRF) of the matter, and Latin indices such as $i \in \{x,y,z\}$ are spatial indices.
 
Assuming a gas of massive quasiparticles, the relaxation time is given by $\tau_{\rm eq}(T)= \bar{\eta} \, (\epsilon+P)/I_{3,2}(m/T)$,
where $\bar\eta = \eta/s$ is the specific shear viscosity, $\epsilon$ is the energy density, and $P$ is the pressure~\cite{Nopoush:2015yga}. The special function $I_{3,2}$ is defined in Ref.~\cite{Nopoush:2015yga}.
The effective temperature $T(\tau)$ is determined by requiring that the nonequilibrium kinetic energy density calculated from $f$ be equal to the equilibrium kinetic energy density calculated from $f_\text{eq}(T,m)$.

We assume the nonequilibrium distribution function is given by the leading-order aHydroQP form, parametrized by a diagonal anisotropy tensor in the fluid LRF
\bqa
f_{\rm LRF}(x,p) &=& f_{\rm eq}\!\left(\frac{1}{\lambda}\sqrt{\sum_i \frac{p_i^2}{\alpha_i^2} + m^2}\right) .
\label{eq:fdef}
\eqa
As indicated above, in the LRF, the argument of the distribution function can be expressed in terms of three independent momentum-anisotropy parameters $\alpha_i$ and a scale parameter $\lambda$, which are space-time dependent fields.  We assume that $f_{\rm eq}$ is given by a Boltzmann distribution.

To determine the space-time evolution of the fields $\vec{u}$, $\vec{\alpha}$, and $\lambda$ one must obtain and then solve seven dynamical equations. The first aHydroQP equation of motion is obtained from the first moment of the left-hand side of the quasiparticle Boltzmann equation, Eq.~\eqref{eq:be}, which reduces to $\partial_\mu T^{\mu\nu}$. In the relaxation time approximation, however, the first moment of the collision kernel on the right-hand side results in a constraint that must be satisfied in order to conserve energy and momentum. This constraint can be enforced by expressing the effective temperature in terms of the microscopic LRF parameters $\lambda$ and $\vec\alpha$.  As a consequence, computing the first moment of the Boltzmann equation gives energy-momentum conservation, $\partial_\mu T^{\mu\nu}=0$.

The second equation of motion required is obtained from the second moment of the quasiparticle Boltzmann equation~\cite{Tinti:2013vba,Nopoush:2015yga}  
\beq
\partial_\alpha  I^{\alpha\nu\lambda} - J^{(\nu} \partial^{\lambda)} m^2 =-\int \dd P \, p^\nu p^\lambda \, {\cal C}[f]\, \label{eq:I-conservation} ,
\eeq
with $I^{\mu\nu\lambda} = \int \dd P \, p^\mu p^\nu p^\lambda f $ and $J^{\mu} = \int \dd P \, p^\mu f$, where $\int \dd P = \int \dd^3\vec{p}/((2\pi)^3 E)$.

%%%%%%%%%%%%%%%%%%%%%%%%%%%%%%%%%%%%%%%%%%%%%%%%%%%%%%%%%%%%%%%%
\subsubsection{The equation of state for aHydroQP}
\label{sec:EoS}
%%%%%%%%%%%%%%%%%%%%%%%%%%%%%%%%%%%%%%%%%%%%%%%%%%%%%%%%%%%%%%%%

The equilibrium kinetic (kin) energy density, pressure, and entropy density for a system of constant-mass particles obeying Boltzmann statistics are given by 
\bqa
\epsilon_{\rm kin}(T,\hat{m}_{\rm eq}) &=& \hat{N} T^4 \, \hat{m}_{\rm eq}^2
 \Big[ 3 K_{2}\left( \hat{m}_{\rm eq} \right) + \hat{m}_{\rm eq} K_{1} \left( \hat{m}_{\rm eq} \right) \Big] , \nonumber \\
 P_{\rm kin}(T,\hat{m}_{\rm eq}) &=& \hat{N} T^4 \, \hat{m}_{\rm eq}^2 K_2\left( \hat{m}_{\rm eq}\right) ,  \nonumber  \\
 s_{\rm kin}(T,\hat{m}_{\rm eq}) &=& \hat{N} T^3 \, \hat{m}_{\rm eq}^2 \Big[4K_2\left( \hat{m}_{\rm eq}\right)+\hat{m}_{\rm eq}K_1\left( \hat{m}_{\rm eq}\right)\Big] , \nonumber  \\
\label{eq:SeqConstantM}
\eqa
where $\hat{m}_{\rm eq} = m/T$, $K_i$ are modified Bessel functions of the second kind, and $\hat{N} = N_{\rm dof}/{(2\pi^2)}$, where $N_{\rm dof}$ is the effective number of degrees of freedom present in the theory under consideration.

In the quasiparticle approach, one assumes that the mass is temperature dependent. This results in a change in the bulk variables in Eqs.~\eqref{eq:SeqConstantM}. When the mass of the quasiparticle depends on the temperature, one cannot simply insert $m(T)$ into the bulk variables, since this will not be thermodynamically consistent.   This is because the entropy density can be obtained in two ways: $s_{\rm eq} = (\epsilon_{\rm eq} + P_{\rm eq})/T$ and $ s_{\rm eq} =\partial P_{\rm eq}/\partial T$. In order to guarantee that both result in the same expression for the entropy density, the energy-momentum tensor definition must include a background field, {\it i.e.}, 
$T^{\mu\nu} = T^{\mu\nu}_{\rm kin} + g^{\mu\nu} B$, where $B$ is an additional nonequilibrium background contribution. 

As a result, in an equilibrium Boltzmann gas of massive quasiparticles, the bulk thermodynamic variables for the gas become
\bqa
\epsilon_{\rm eq}(T,\hat{m}_{\rm eq}) &=& \epsilon_{\rm kin} +B_{\rm eq} \, , 
\label{eq:Eeq} \nonumber \\
 P_{\rm eq}(T,\hat{m}_{\rm eq}) &=& P_{\rm kin} -B_{\rm eq} \, , \nonumber \\
 \label{eq:Peq}
 s_{\rm eq}(T,\hat{m}_{\rm eq}) &=&s_{\rm kin} \, ,
\label{eq:Seq}
\eqa
where $B_{\rm eq}$ is the equilibrium limit of $B$.

To determine the temperature dependence of $B_{\rm eq}$ we require thermodynamic consistency,
\beq
\epsilon_{\rm eq} + P_{\rm eq} = T \frac{\partial P_{\rm eq}}{\partial T} \, ,
\eeq
which provides a first-order differential equation that can be used to determine $B_{\rm eq}(T)$ once $m(T)$ is specified~\cite{Alqahtani:2015qja}.

The quasiparticle mass itself can be determined from the entropy density since it is independent of $B_{\rm eq}$,
\beq
\epsilon_{\rm eq} + P_{\rm eq} = T s_{\rm eq} =  \hat{N} T^4 \, \hat{m}_{\rm eq}^3 K_3\left( \hat{m}_{\rm eq}\right) .
\eeq
In practice, one can determine $m(T)$ from the sum of the equilibrium energy density and pressure determined from lattice QCD calculations~\cite{Alqahtani:2015qja}. The resulting effective mass, scaled by $T$, extracted from the continuum extrapolated Wuppertal-Budapest lattice data \cite{Borsanyi:2010cj} can be found in Refs.~\cite{Alqahtani:2015qja,Ryblewski:2017ybw}.   At high temperatures, \mbox{$T \gtrsim$ 500 MeV}, the mass is proportional to $T$, in agreement with the expected high-temperature behavior of QCD quasiparticles and, at low temperatures, $T \lesssim 100$ MeV, the extracted mass is consistent with a pion gas~\cite{Ryblewski:2017ybw}.  Plots of $m(T)$ and $B_{\rm eq}(T)$ can be found in Fig.~2 of Ref.~\cite{Alqahtani:2015qja}.

%%%%%%%%%%%%%%%%%%%%%%%%%%%%%%%%%%%%%%%%%%%%%%%%%%%%%%%%%%%%%%%%
\subsubsection{Evolution equations in aHydroQP}
\label{sec:ahydroeqs}
%%%%%%%%%%%%%%%%%%%%%%%%%%%%%%%%%%%%%%%%%%%%%%%%%%%%%%%%%%%%%%%%

The evolution equations for $\vec{u}$, $\lambda$, and $\vec{\alpha}$ are obtained from moments of the quasiparticle Boltzmann equation.  These can be expressed compactly by introducing a timelike vector $u^\mu$, which is normalized as $u^\mu u_\mu = 1$ and three spacelike vectors $X_i^\mu$, which are individually normalized as $X^\mu_i X_{\mu,i} = -1$~\cite{Ryblewski:2010ch,Martinez:2012tu}.  These vectors are mutually orthogonal and obey $u_\mu X^\mu_i = 0$ and $X_{\mu,i} X^\mu_j = 0$ for $i \neq j$.  The four equations resulting from the first moment of the quasiparticle Boltzmann equation are
\bqa
&& D_u\epsilon +\epsilon \theta_u + \sum_j P_j u_\mu D_j X^\mu_j = 0\, , \label{eq:1stmomOne} \\
&& D_i P_i+P_i\theta_i -\epsilon X_{\mu,i} D_uu^\mu + P_i X_{\mu,i} D_i X^\mu_i \nonumber \\
&& \hspace{3cm} - \sum_j P_j X_{\mu,i} D_j X^\mu_j  = 0\,, 
\label{eq:1stmomTwo} 
\eqa
where $D_u \equiv u^\mu \partial_\mu$ and $D_i \equiv X^\mu_i \partial_\mu$.  The expansion scalars are $\theta_u = \partial_\mu u^\mu$ and $\theta_i = \partial_\mu X^\mu_i$.  Explicit expressions for all basis vectors, derivative operators, and expansion scalars can be found in Refs.~\cite{Nopoush:2014pfa,Alqahtani:2015qja,Alqahtani:2016rth,Alqahtani:2017tnq}.  The quantities $\epsilon$ and $P_i$ are the kinetic energy density and pressures obtained using the anisotropic hydrodynamics ansatz for the one-particle distributions function corrected by the background contribution $B$ necessary to enforce thermodynamic consistency,
\bqa
\epsilon &=& \epsilon_{\rm kin}(\lambda,\vec\alpha,m) + B(\lambda,\vec\alpha) \, ,  \\
P_i &=& P_{i, \rm kin}(\lambda,\vec\alpha,m) - B(\lambda,\vec\alpha) \, .
\eqa

The three equations resulting from the second moment of the Boltzmann equation are
\beq
D_u I_i + I_i (\theta_u + 2 u_\mu D_i X_i^\mu)
= \frac{1}{\tau_{\rm eq}} \Big[ I_{\rm eq}(T,m) - I_i \Big] ,
\label{eq:2ndmoment} 
\eeq
where $I_i = u^\mu X^\nu_i X^\lambda_i I_{\mu\nu\lambda} = \alpha \, \alpha_i^2 \, I_{\rm eq}(\lambda,m)$ with $I_{\rm eq}(\lambda,m) = \hat{N} \lambda^2 m^3 K_3(m/\lambda)$ and \mbox{$\alpha = \prod_j \alpha_j$}~\cite{Nopoush:2014pfa}.

Equations~\eqref{eq:1stmomOne}, \eqref{eq:1stmomTwo}, and \eqref{eq:2ndmoment} provide seven partial differential equations for $\vec{u}$, $\vec{\alpha}$, and $\lambda$ which we solve numerically.  To determine the local effective temperature, we make use of the constraint requiring that the equilibrium and nonequilibrium energy densities in the LRF be equal (Landau matching).  The resulting system of partial differential equations is evolved until the effective temperature in the entire simulation volume falls below a given freeze-out temperature, $T_{\rm fo}$.  

%%%%%%%%%%%%%%%%%%%%%%%%%%%%%%%%%%%%%%%%%%%%%%%%%%%%%%%%%%%%%%%%
\subsubsection{Initial transverse and longitudinal profiles}
%%%%%%%%%%%%%%%%%%%%%%%%%%%%%%%%%%%%%%%%%%%%%%%%%%%%%%%%%%%%%%%%

The initial distribution of the energy density in the transverse plane is computed from a ``tilted'' profile \cite{Bozek:2013uha,Bozek:2010bi,Bzdak:2009xq,Bialas:2004su}.  The distribution used is a linear combination of optical Glauber wounded nucleon and binary collision density profiles, with a binary collision mixing factor of $\chi = 0.15$ taken from previous studies~\cite{Alqahtani:2017jwl,Alqahtani:2017tnq,Alqahtani:2020paa}.  We assume that the inelastic nucleon-nucleon scattering cross section is \mbox{67.6 mb} and \mbox{71 mb} at $\sqrt{s_{NN}} = 5.02$ and 8.16 TeV, respectively. We use an impact parameter of \mbox{$b = 4.70$ fm} and \mbox{4.71 fm} at $\sqrt{s_{NN}} =$ 5.02 and \mbox{8.16 TeV}, respectively.  Both values for the impact parameter were obtained by determining the min-bias impact parameter in the optical Glauber model.  The corresponding number of participant nucleons in min-bias collisions is 9.45 and 9.98, respectively.  In the longitudinal direction, we use a profile with a tilted central plateau and Gaussian tails in the fragmentation region, resulting in a longitudinal profile function of the form
\beq
\rho(\varsigma) \equiv \exp \left[ - (\varsigma - \Delta \varsigma)^2/(2 \sigma_\varsigma^2) \, \Theta (|\varsigma| - \Delta \varsigma) \right] ,
\label{eq:rhofunc}
\eeq
where $\varsigma = {\rm arctanh}(z/t)$ is the spatial rapidity.  We fit the central width $\Delta\varsigma$ to the pseudorapidity distribution of charged hadron production, finding \mbox{$\Delta\varsigma = 1.8$} for $\sqrt{s_{NN}} =$ 5.02 TeV collisions and $\Delta\varsigma = 2.1$ for \mbox{$\sqrt{s_{NN}} =$ 8.16 TeV} collisions.  We note that the width of the Gaussian tails, $\sigma_{\varsigma}$, is largely unconstrained based on available $p$-Pb data.  As a result, we have used $\sigma_{\varsigma} = 1.6$, which was used previously in Pb-Pb collisions at both $\sqrt{s_{NN}} =$ 5.02 TeV and 2.76 TeV~\cite{Alqahtani:2017jwl,Alqahtani:2017tnq,Alqahtani:2020paa}.

The resulting initial energy density at a given transverse position ${\vec x}_\perp$ and spatial rapidity $\varsigma$ was computed using~\cite{Bozek:2013uha} 
\bqa
\epsilon({\vec x}_\perp,\varsigma) &\propto& (1-\chi) \rho(\varsigma) \Big[ W_p({\vec x}_\perp) g(\varsigma) + W_A({\vec x}_\perp) g(-\varsigma)\Big] \nonumber \\
&& \hspace{3cm} + \; \chi \rho(\varsigma) C({\vec x}_\perp) \, ,
\eqa
where $W_A({\vec x}_\perp)$ is the wounded-nucleon density for nucleus $A$ \cite{Florkowski2010-tl}, $C({\vec x}_\perp)$ is the binary collision density~\cite{Florkowski2010-tl}, and $g(\varsigma)$ is the tilt function.  To compute $W_p$, we parametrize the proton overlap function as~\cite{Skands:2014pea,dEnterria:2020dwq}
\beq
T_p({\vec b})= \frac{n}{2\pi r^2_p \,\Gamma (2/n)} \exp{[-(b/r_p)^n]}\,,
\label{eq:overlap}
\eeq
with $n = 1.85$ and $r_p = 0.975$ fm.

The tilt function is defined as~\cite{Bozek:2013uha}
\bqa 
g(\varsigma) =
\left\{ \begin{array}{lcc}
0  &  \,\,\, & \varsigma < -y_N \, , \\ 
(\varsigma+y_N)/(2y_N) & & -y_N \leq \varsigma \leq y_N \, , \\
1 & & \varsigma > y_N \, ,
\end{array}\right. 
\eqa 
where $y_N = \log(2\sqrt{s_{NN}}/(m_p + m_n))$ is the nucleon rapidity.

%%%%%%%%%%%%%%%%%%%%%%%%%%%%%%%%%%%%%%%%%%%%%%%%%%%%%%%%%%%%%%%%
\subsubsection{Freeze-out and hadronic decays}
%%%%%%%%%%%%%%%%%%%%%%%%%%%%%%%%%%%%%%%%%%%%%%%%%%%%%%%%%%%%%%%%

We extract a three-dimensional freezeout hypersurface at fixed energy density (temperature) corresponding to $T_{\rm fo}$ = 130 MeV from the aHydroQP evolution. We assume that the fluid anisotropies $\vec\alpha$ and scale parameter $\lambda$ are the same for all hadron species.  We also assume that all hadrons are created in chemical equilibrium. With the use of an extended Cooper-Frye prescription \cite{Alqahtani:2017mhy}, the underlying hydrodynamic values for the flow velocity, the anisotropy parameters, and scale parameter can be translated into primordial hadron distributions on this hypersurface. 

The values of the aHydroQP parameters on the freezeout hypersurface are processed by a modified version of {\sc Therminator 2} that generates hadronic configurations using Monte Carlo sampling~\cite{Chojnacki:2011hb}. After sampling the primordial hadrons, hadronic decays are taken into account using the built-in routines in {\sc Therminator 2}. The source code for aHydroQP and the custom version of {\sc Therminator 2} used are publicly available~\cite{kent-code-library}.

%-----------------------------------------------
\begin{figure}[t]
\begin{center}
\includegraphics[width=0.95\linewidth]{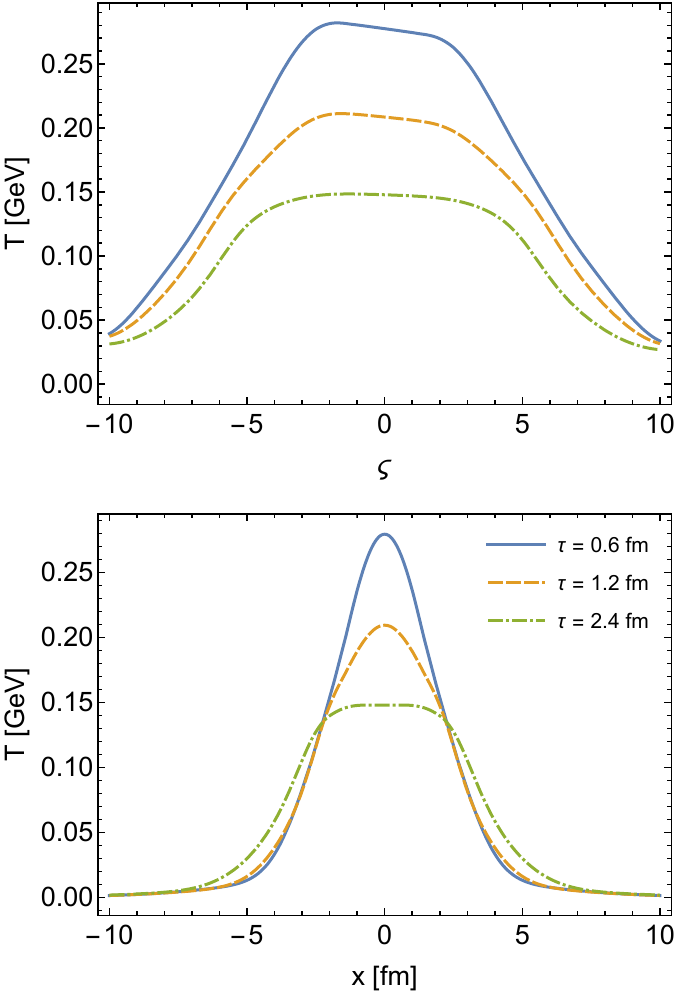}
\vspace{-5mm}
\end{center}
\caption{Evolution of the temperature obtained using aHydroQP for $\sqrt{s_{NN}} = $ 5.02 TeV $p$-Pb collisions.  The top panel shows the temperature as a function of spatial rapidity $\varsigma$ and the bottom panel shows the temperature as a function of the transverse position $x$.  In both panels, the solid, dashed, and dot-dashed curves show the temperature profile at $\tau = $ 0.6 fm, 1.2 fm, and 2.4 fm, respectively.} 
\label{fig:evolPlot-5TeV}
\end{figure}
%-----------------------------------------------

%-----------------------------------------------
\begin{figure}[t]
\begin{center}
\includegraphics[width=1\linewidth]{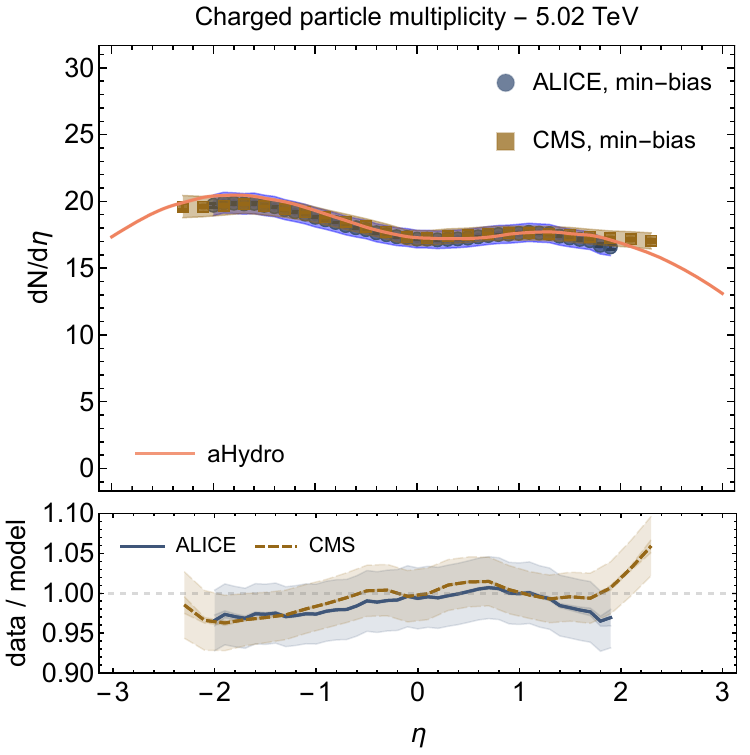}
\end{center}
\vspace{-5mm}
\caption{Min-bias charged particle multiplicity $dN/d\eta$ for $\sqrt{s_{NN}} = $ 5.02 TeV $p$-Pb collisions.
ALICE and CMS Collaboration data are from Refs.~\cite{ALICE:2012xs} and \cite{CMS:2017shj}, respectively.  The top panel compares the results of the aHydroQP model to the experimental data while the bottom panel shows the relative error.  In both panels, the shaded regions indicate the reported experimental uncertainty.}
\label{fig:dndeta-5TeV}
\end{figure}
%-----------------------------------------------

%-----------------------------------------------
\begin{figure}[t]
\begin{center}
\includegraphics[width=1\linewidth]{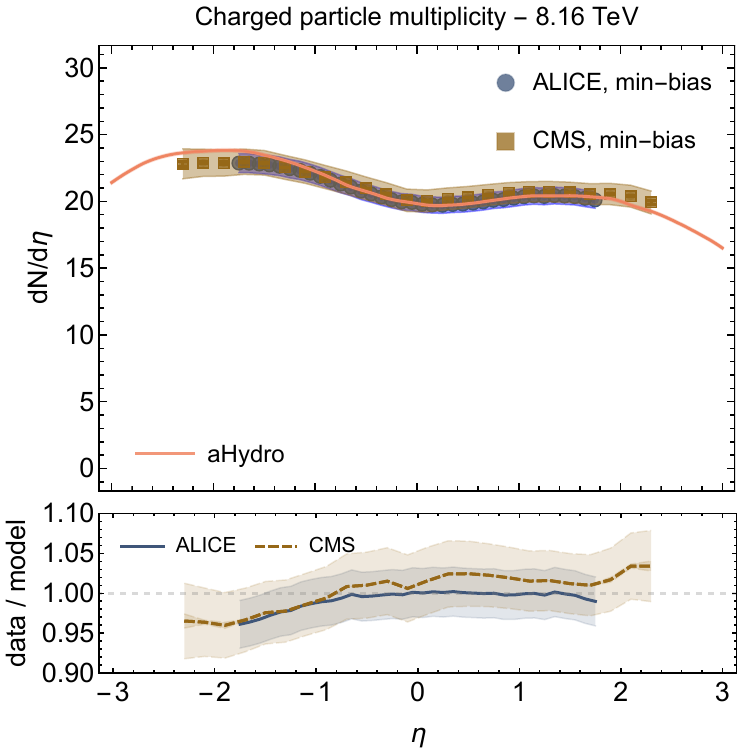}
\end{center}
\vspace{-5mm}
\caption{Min-bias charged particle multiplicity $dN/d\eta$ for $\sqrt{s_{NN}} = $ 8.16 TeV $p$-Pb collisions. ALICE and CMS Collaboration data are from Refs.~\cite{ALICE:2018wma} and \cite{CMS:2017shj}, respectively.  The panels are the same as in Fig.~\ref{fig:dndeta-5TeV}.} 
\label{fig:dndeta-8TeV}
\end{figure}
%-----------------------------------------------

%-----------------------------------------------
\begin{figure*}[t]
\begin{center}
\includegraphics[width=0.975\linewidth]{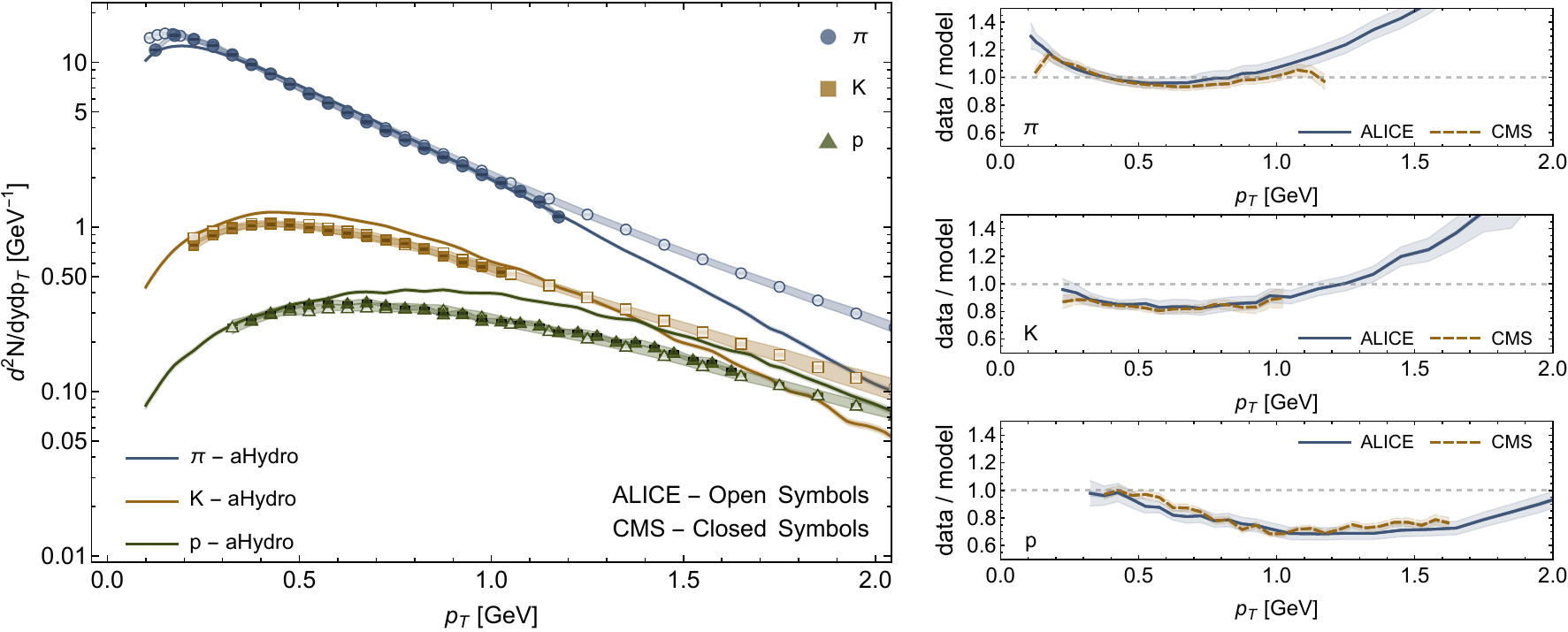}
\end{center}
\vspace{-5mm}
\caption{Min-bias identified particle spectra of pions, kaons, and protons in $p$-Pb collisions at $\sqrt{s_{NN}} = $ 5.02 TeV compared to aHydroQP predictions.  Experimental data from the ALICE and CMS Collaborations are from Refs.~\cite{ALICE:2013wgn} and \cite{CMS:2013pdl}, respectively.  The left panel shows model comparisons to the data and the three right panels show the relative error.  In both panels, the shaded regions indicate the reported experimental uncertainty.} 
\label{fig:spectra-5TeV}
\end{figure*}
%-----------------------------------------------

%%%%%%%%%%%%%%%%%%%%%%%%%%%%%%%%%%%%%%%%%%%%%%%%%%%%%%%%%%%%%%%%
\subsubsection{Comparisons with experimental data}
%%%%%%%%%%%%%%%%%%%%%%%%%%%%%%%%%%%%%%%%%%%%%%%%%%%%%%%%%%%%%%%%

Previously, the aHydroQP formalism has been used to describe Pb-Pb collisions at the LHC and Au-Au collisions at RHIC for energies of 2.76 TeV, 5.02 TeV, and 200 GeV, respectively. It was found that the observed spectra of identified hadrons, charged particle multiplicities, elliptic flow, Hanbury-Brown-Twiss radii, etc. were well reproduced~\cite{Alqahtani:2017jwl,Alqahtani:2017tnq,Almaalol:2018gjh,Alqahtani:2020paa,Alqahtani:2020daq,Alqahtani:2020paa,Alqahtani:2022erx}. Additionally, we note that aHydroQP has been used as the hydrodynamic background for many previous calculations of bottomonium suppression and elliptic flow in $AA$ collisions \cite{Bhaduri:2018iwr,Bhaduri:2020lur,Islam:2020gdv,Islam:2020bnp,Brambilla:2020qwo,Brambilla:2021wkt,Brambilla:2022ynh,Brambilla:2023hkw,Strickland:2023nfm}.

Here we apply aHydroQP to min-bias $p$-Pb collisions at both $\sqrt{s_{NN}} =$ 5.02 TeV and 8.16 TeV to use it as a hydrodynamic background in the calculation of QGP-induced bottomonium suppression.  The same simulation volume was used for both energies: $L_\perp = 32$ fm in the transverse direction and 20 units of spatial rapidity in the longitudinal direction.  The box size is taken to be rather large to ensure that our results are not affected by the boundaries.  We used 80 grid points in each of the transverse directions and 33 grid points in the longitudinal direction.  The coordinate system for the simulation was centered on the proton in the transverse plane, with the Pb nucleus shifted to the left by the min-bias impact parameter appropriate for each collision energy.

An initialization time of \mbox{$\tau_0 = 0.25$~fm/$c$} was used for both collision energies.  At this proper time, the initial transverse momentum anisotropies were taken to be \mbox{$\alpha_{x,0} = \alpha_{y,0} = 1$}, and the initial longitudinal momentum anisotropy was taken to be $\alpha_{z,0} = 0.2$.  The initial longitudinal momentum anisotropy used was chosen to reflect the high degree of early-time momentum anisotropy in the fluid LRF~\cite{Strickland:2013uga}.  We assumed that the initial fluid flow velocity was zero in the transverse directions and boost-invariant Bjorken flow in the longitudinal direction.  The best fit initial temperatures at \mbox{$\tau_0 = 0.25$~fm/$c$}, $\vec{x}=0$, and $\varsigma=0$ are \mbox{$T_0 = $ 480 MeV} at \mbox{$\sqrt{s_{NN}} =$ 5.02 TeV} and $T_0 = $ 496 MeV at \mbox{$\sqrt{s_{NN}} =$ 8.16 TeV}.  In the simulations performed, we assumed a temperature-independent shear viscosity to entropy density ratio, with the temperature dependence of the bulk viscosity determined self-consistently using the massive quasiparticle model~\cite{Alqahtani:2017jwl,Alqahtani:2017tnq}.  Our best fits to the charged particle multiplicities and identified particle spectra were obtained with $\eta/s = 0.32$ at both collision energies.

In Fig.~\ref{fig:evolPlot-5TeV} we present the temperature evolution of the aHydroQP runs used for the $p$-Pb background at $\sqrt{s_{NN}} = $ 5.02 TeV.  The top and bottom panels show the spatial rapidity, $\varsigma$, and transverse coordinate, $x$, dependence of the temperature obtained from the aHydroQP evolution.  As can be seen from the top panel of Fig.~\ref{fig:evolPlot-5TeV}, the tilted initial conditions result in a temperature profile that is peaked on the Pb-going side (negative spatial rapidity).  This panel also shows how the temperature decreases at large spatial rapidity.  As we show below, this decrease in temperature at high rapidity results in a small QGP-induced suppression at large rapidity.  In addition, in the central rapidity region, $|y| \lesssim 3$, we find that the tilted initial conditions lead to stronger QGP-induced suppression on the Pb-going side than on the $p$-going side.  Finally, we note that due to the lower initial temperature compared to Pb-Pb collisions at \mbox{$\sqrt{s_{NN}} = $ 5.02 TeV}, on average, the lifetime of the QGP generated in $p$-Pb collisions is shorter than that in Pb-Pb collisions where the min-bias lifetime is on the order of 10 fm.  This shorter lifetime results in much smaller QGP-induced bottomonium suppression than is found in Pb-Pb collisions at the same $\sqrt{s_{NN}}$.

To assess our tuning of the aHydroQP $p$-Pb background evolution, in Figs.~\ref{fig:dndeta-5TeV}-\ref{fig:spectra-5TeV} we present comparisons of our aHydroQP $p$-Pb results with standard soft hadron observables measured at LHC energies.  In Figs.~\ref{fig:dndeta-5TeV} and \ref{fig:dndeta-8TeV} we present our aHydroQP results for the charged particle multiplicity as a function of the pseudorapidity, $\eta \equiv \tfrac{1}{2} \ln((p+p_z)/(p-p_z))$, at $\sqrt{s_{NN}} = $ 5.02 TeV and $\sqrt{s_{NN}} = $ 8.16 TeV, respectively.  In these figures, we compare to data from the ALICE \cite{ALICE:2012xs,ALICE:2013wgn} and CMS \cite{CMS:2013pdl} Collaborations.  In both figures, the top panel shows the model results compared to ALICE and CMS data while the bottom panel shows the relative error (data/model) of the aHydroQP results.  As can be seen from these figures, our aHydroQP evolution reproduces the measurements of the ALICE and CMS Collaborations to within approximately 5\% in the reported pseudorapidity range.  This level of agreement is comparable to or better than all model comparisons shown in Refs.~\cite{ALICE:2012xs,ALICE:2013wgn,CMS:2013pdl}.  

In order to demonstrate the predictive power of the aHydroQP model, in Fig.~\ref{fig:spectra-5TeV} we present comparisons of the computed transverse momentum spectra in $p$-Pb collisions at $\sqrt{s_{NN}} = $ 5.02 TeV for pions, kaons, and protons with data from the ALICE \cite{ALICE:2013wgn} and CMS Collaborations \cite{CMS:2013pdl}.  In the left panel, we show comparisons of the identified spectra obtained using aHydroQP with experimental data and in the three right panels we plot the relative error (data/model).  As can be seen from this figure, aHydroQP can reproduce the experimental observations for $p_T \lesssim 1.5$ GeV reasonably well.  The differences that remain are comparable to those found using the other models presented in Refs.~\cite{ALICE:2013wgn,CMS:2013pdl}, providing confidence that the aHydroQP background provides a reasonable background for computing QGP-induced bottomonium suppression.  

We note that at high transverse momentum, it is expected that hard processes play an important role, and as a result, viscous hydrodynamics is not expected to accurately reproduce the data.  Again, we emphasize that the agreement of our aHydroQP model predictions with available data is comparable to other state-of-the-art models in the literature.  Since pions represent the vast majority of produced particles, our agreement with $dN/d\eta$ and the pion spectra implies that the hydrodynamical background is sufficiently accurate to estimate the effect of QGP-induced bottomonium suppression in $p$-Pb collisions.

%%%%%%%%%%%%%%%%%%%%%%%%%%%%%%%%%%%%%%%%%%%%%%%%%%%%%%%%%%%%%%%%
\subsection{Coupling bottomonium suppression to aHydroQP}
%%%%%%%%%%%%%%%%%%%%%%%%%%%%%%%%%%%%%%%%%%%%%%%%%%%%%%%%%%%%%%%%

Having obtained good agreement with soft-hadron observables, we now turn to the use of the generated aHydroQP evolution as the hydrodynamic background for computing bottomonium suppression.  The evolution of the hydrodynamic background is carried out in Milne coordinates $\tilde{x}^\mu = (\tau, \vec{x}_\perp, \varsigma)$ instead of Minkowski coordinate $ x^\mu = (t, \vec{x}_\perp, z)$, with longitudinal proper time $\tau = \sqrt{t^2-z^2}$ and space-time rapidity $\varsigma = {\rm arctanh}(z/t)$. Therefore, we need to express bottomonium spacetime positions in the Milne coordinates through the transformation $t = \tau \cosh{\varsigma}$, $z = \tau \sinh{\varsigma}$ and $v_z = \tanh{y}$, where $y$ is the momentum rapidity and $v_z$ is the velocity of the particle along the beam axis. Here, we assume that the bottomonium states travel along eikonal trajectories after their initial production. In Minkowski coordinates, we assume that the bottomonium states move at a constant velocity such that $\vec{x} = \vec{x}_{0} + \vec{v} (t - t_{0})$, where $\vec{x}_0$ is the position of the state at $t_0$ and $\vec{v} = \vec{p}_0/E$ with $E = \sqrt{\vec{p}^2 + M_\Upsilon^2}$. Expressing this in terms of Milne coordinates, we obtain
\bqa
\vec{x}_\perp &=& \vec{x}_{\perp,0} + \vec{v}_\perp (\tau \cosh{\varsigma} - \tau_{0} \cosh{\varsigma_{0}}) \, ,  \nonumber \\
\tau \sinh{\varsigma} &=& \tau_{0} \sinh{\varsigma_{0}} + \tanh{y} \, (\tau \cosh{\varsigma} - \tau_{0} \cosh{\varsigma_{0}}) \, , \nonumber \\
\label{eq:milne}
\eqa
where $\varsigma_0$ is the spatial rapidity at initial proper time $\tau_0$ and $\vec{v}_\perp$ is the transverse velocity. Using these equations, one can obtain the Milne coordinates of a bottomonium state based on its initially sampled momentum and position. Equation ~\eqref{eq:milne} can be solved for $\varsigma$ giving $\varsigma = y + \log({\cal G})$ with 
\beq
{\cal G} \equiv \sqrt{1 + \frac{\sinh^2(\varsigma_0-y)}{\bar\tau^2}} + \frac{\sinh(\varsigma_0-y)}{\bar\tau}  \, ,
\eeq
and $\bar\tau \equiv \tau/\tau_0$. As $\bar\tau \rightarrow \infty$, ${\cal G} \rightarrow 1$, and when $\bar\tau \rightarrow 1$, one obtains ${\cal G} \rightarrow \exp(\varsigma_0-y)$. Note that $\varsigma=\varsigma_0 = y$ at all times in the limit that all production occurs at $\tau_0 \rightarrow 0$.  In the limit $M_\Upsilon \rightarrow \infty$, all bottomonium production occurs at $\tau_0=0$.

Here, we sample 3D trajectories for bottomonium states and use the background temperature evolution generated by the tuned 3+1D aHydroQP runs. The initial transverse positions for the bottomonium production in $p$-Pb collisions are Monte Carlo sampled on the basis of the binary collision overlap profile of the proton and Pb nucleus calculated from the Glauber model. We sample the initial transverse momenta and momentum rapidities from a distribution of the form given in Eq.~\eqref{eq:ppdist} using the average mass of the $\Upsilon(1S)$, $\Upsilon(2S)$, and $\Upsilon(3S)$ states.  Finally, we sample the initial azimuthal angle $\phi$ of the produced bottomonium uniformly in the range $[0,2\pi)$.

We use NLO QTraj \cite{qtraj-download} to simulate the quantum dynamics of each sampled physical trajectory for the bottomonium states in the QGP. 
 After the evolution is complete, the survival probabilities of each bottomonium state are computed as the ratio of the modulus squared of each eigenstate's overlaps with the final and initial wave functions, respectively. The final observables are then averaged over a large set of physical trajectories.

The QTraj calculations use a one-dimensional lattice size $L = 40$ GeV$^{-1}$ with 2048 points.  In our calculations, we averaged over 160,000 physical trajectories such that the statistical error was much smaller than the systematic error resulting from varying the heavy quarkonium transport coefficient $\hat\kappa$.  We used a decoupling temperature of $T_F = $ 180 MeV.  We varied the heavy quarkonium momentum diffusion constant, $\hat\kappa$, in the range $\hat\kappa \in \{5,6,7\}$ and assumed that $\hat\gamma = 0$, consistent with recent observations that the bottomonia mass shift in the QGP is very small \cite{Larsen:2019bwy}.  This latter assumption is also consistent with Ref.~\cite{Brambilla:2023hkw} where $\hat\gamma = 0$ was found to give the best description of the available ground- and excited-state bottomonium suppression data in \mbox{$\sqrt{s_{NN}} = $ 5.02 TeV} Pb-Pb collisions at the LHC.  Due to the computational cost, we ignored the effect of quantum jumps (quantum regeneration) since this effect is subleading at small $N_{\rm part}$.  We have verified this statement by explicit calculations that include quantum jumps in the case of min-bias $p$-Pb collisions and found that the corrections coming from jumps were on the 2\% level for all states and therefore subleading compared to the uncertainties associated with varying the heavy quarkonium transport coefficient $\hat\kappa$.

%-----------------------------------------------
\begin{figure}[t]
\begin{center}
\includegraphics[width=0.975\linewidth]{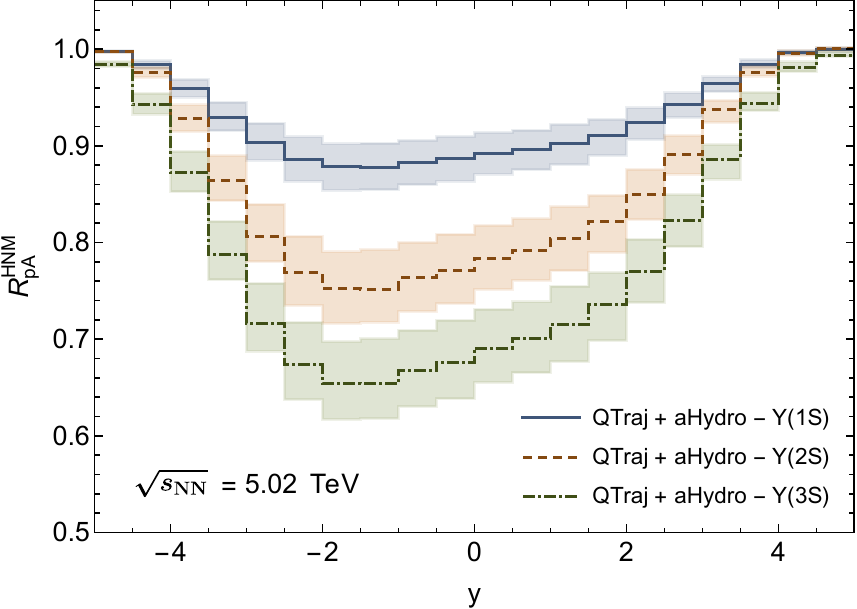}
\end{center}
\vspace{-5mm}
\caption{Hot QGP induced $\Upsilon(1S)$ (solid blue), $\Upsilon(2S)$ (dashed orange), and $\Upsilon(3S)$ (long dashed green) $R_{pA}^{\rm HNM}$ as a function of $y$ at $\sqrt{s_{NN}} = $ 5.02 TeV.  The shaded bands correspond to the theoretical uncertainty associated with varying $\hat\kappa \in \{5,6,7\}$.} 
\label{fig:rpA-y-qtraj-5.02}
\end{figure}
%-----------------------------------------------

In Fig.~\ref{fig:rpA-y-qtraj-5.02} we present our results for the suppression obtained by folding the QTraj evolution of the quantum wave function together with our tuned \mbox{$\sqrt{s_{NN}} = $ 5.02 TeV} aHydroQP hydrodynamical background.  The shaded bands indicate the results obtained when varying our QTraj model parameters in the ranges described above.  As this figure demonstrates, we find that all states experience increased suppression on the Pb-going side ($y < 0$) because the temperature on the Pb-going side is higher, as shown in Fig.~\ref{fig:evolPlot-5TeV}.  We also see that the excited states experience stronger suppression with $R_{pA}^{\rm HNM}(1S) > R_{pA}^{\rm HNM}(2S) > R_{pA}^{\rm HNM}(3S)$, because the decay widths of the excited states are ordered with $\Gamma(1S) < \Gamma(2S) < \Gamma(3S)$.  Finally, we see that, when the magnitude of the rapidity is large in either the forward- or backward-going direction, QGP-induced suppression decreases because the temperature decreases rapidly at forward and backward spatial rapidity, as shown in Fig.~\ref{fig:evolPlot-5TeV}.

%%%%%%%%%%%%%%%%%%%%%%%%%%%%%%%%%%%%%%%%%%%%%%%%%%%%%%%%%%%%%%%%
\section{Combining nPDF effects, energy loss, momentum broadening, and final state effects}
\label{sec:combo}
%%%%%%%%%%%%%%%%%%%%%%%%%%%%%%%%%%%%%%%%%%%%%%%%%%%%%%%%%%%%%%%%

Having calculated the bottomonium suppression due to cold and hot nuclear matter effects, we can now combine all effects to obtain the total nuclear modification factor for the quarkonium state in $p$-Pb collisions. We obtain the total quarkonium suppression in $pA$ collisions relative to $pp$ collisions due to cold nuclear matter (CNM) and hot nuclear matter (HNM) as
\beq
R^{\Upsilon}_{pA} = R^{\rm CNM}_{pA}  \times  R^{\rm HNM}_{pA} \, ,
\eeq
where 
\beq
R^{\rm CNM}_{pA} = R^{\rm nPDF}_{pA} 
 \times R^{\rm eloss, broad}_{pA} \, ,
\eeq
is the nuclear modification factor due to cold nuclear matter effects.  Above, $R^{\rm nPDF}_{pA}$ is the suppression due to nPDF effects in $p$-Pb collisions relative to $pp$ collisions as described in Sec.~\ref{sec:npdf} and $R^{\rm eloss, broad}_{pA}$ is the suppression due to coherent energy loss and momentum broadening of bottomonium described in Sec.~\ref{sec:eloss}. The effect of HNM on bottomonium propagation through the QGP is encoded in the factor $R^{\rm HNM}_{pA}$ detailed in Sec.~\ref{sec:eloss}.  Finally, we implement the effect of excited-state feed down as described in the next section.

%%%%%%%%%%%%%%%%%%%%%%%%%%%%%%%%%%%%%%%%%%%%
\section{Excited state feed down}
\label{sec:feeddown}
%%%%%%%%%%%%%%%%%%%%%%%%%%%%%%%%%%%%%%%%%%%%

After emerging from the QGP, the feed down of bottomonium excited states must be taken into account.  We employ a feed-down matrix, denoted as $F$, which establishes an empirical relationship between experimentally observed and directly produced $pp$ cross sections, represented as \mbox{$\vec{\sigma}_{\text{exp}} = F \vec{\sigma}_{\text{direct}}$}.  The vectors $\vec{\sigma}_{\text{direct}}$ and $\vec{\sigma}_{\text{exp}}$ contain the scattering cross sections for the $\Upsilon(1S)$, $\Upsilon(2S)$, $\chi_{b0}(1P)$, $\chi_{b1}(1P)$, $\chi_{b2}(1P)$, $\Upsilon(3S)$, $\chi_{b0}(2P)$, $\chi_{b1}(2P)$, and $\chi_{b2}(2P)$ states before and after feed down, respectively. The feed-down matrix, $F$, is a square matrix with values set by the experimentally determined branching fractions of bottomonium excited states into lower-lying states. In general, the entries in the feed down matrix $F$ are
\beq
F_{ij} = \left\{ \begin{matrix}
\; \text{branching fraction $j$ to $i$},\; & \text{for } i < j \, , \\
1, & \text{for } i = j \, , \\
0, & \text{for } i > j \, ,
\end{matrix} \right.
\eeq
where the branching fractions are taken from the Particle Data Group~\cite{Zyla:2020zbs} (see Appendix of Ref.~\cite{Boyd:2023ybk} for all elements of $F_{ij}$).

The final nuclear modification factor $R_{\rm pA}$ in min-bias $p$-Pb collisions for bottomonium state $i$ is computed using
\beq
R^{i}_{pA}(p_T,y,\phi) = \frac{\left(F \cdot R^{\Upsilon}_{pA}(p_T,y,\phi) \cdot \vec{\sigma}_{\text{direct}}\right)^{i}}{\vec{\sigma}_{\text{exp}}^{i}} \, ,
\label{eq:feeddown}
\eeq
where $R^{\Upsilon}_{pA}(p_T,y,\phi)$ is the total suppression computed from the sequence of initial- and final-state effects described above, $p_T$ is transverse momentum, $y$ is the momentum rapidity, and $\phi$ is the azimuthal angle.  In the results reported below, we integrate over the azimuthal angle and bin $R^{i}_{pA}$ in $p_T$ and $y$.  The $pp$ cross sections are $\vec{\sigma}_{\text{exp}}=\{57.6$, 19, 3.72, 13.69, 16.1, 6.8, 3.27, 12.0, $14.15\}$ nb for \mbox{$\sqrt{s_{NN}} = $ 5.02 TeV} collisions.
These were obtained from the experimental measurements presented in Refs.~\cite{Sirunyan:2018nsz,Aaij:2014caa}, as explained in Sec.~6.4 of Ref.~\cite{Brambilla:2020qwo}.  The cross sections at \mbox{$\sqrt{s_{NN}} = $ 8.16 TeV} are obtained by uniformly scaling the \mbox{$\sqrt{s_{NN}} = $ 5.02 TeV} cross sections by \mbox{8.16/5.02 = 1.63}.  Since the scaling is independent of $p_T$, $y$, $\phi$, and the state under consideration, it cancels in the ratio $R^{i}_{pA}$.

%-----------------------------------------------
\begin{figure}[t]
\begin{center}
\includegraphics[width=0.98\linewidth]{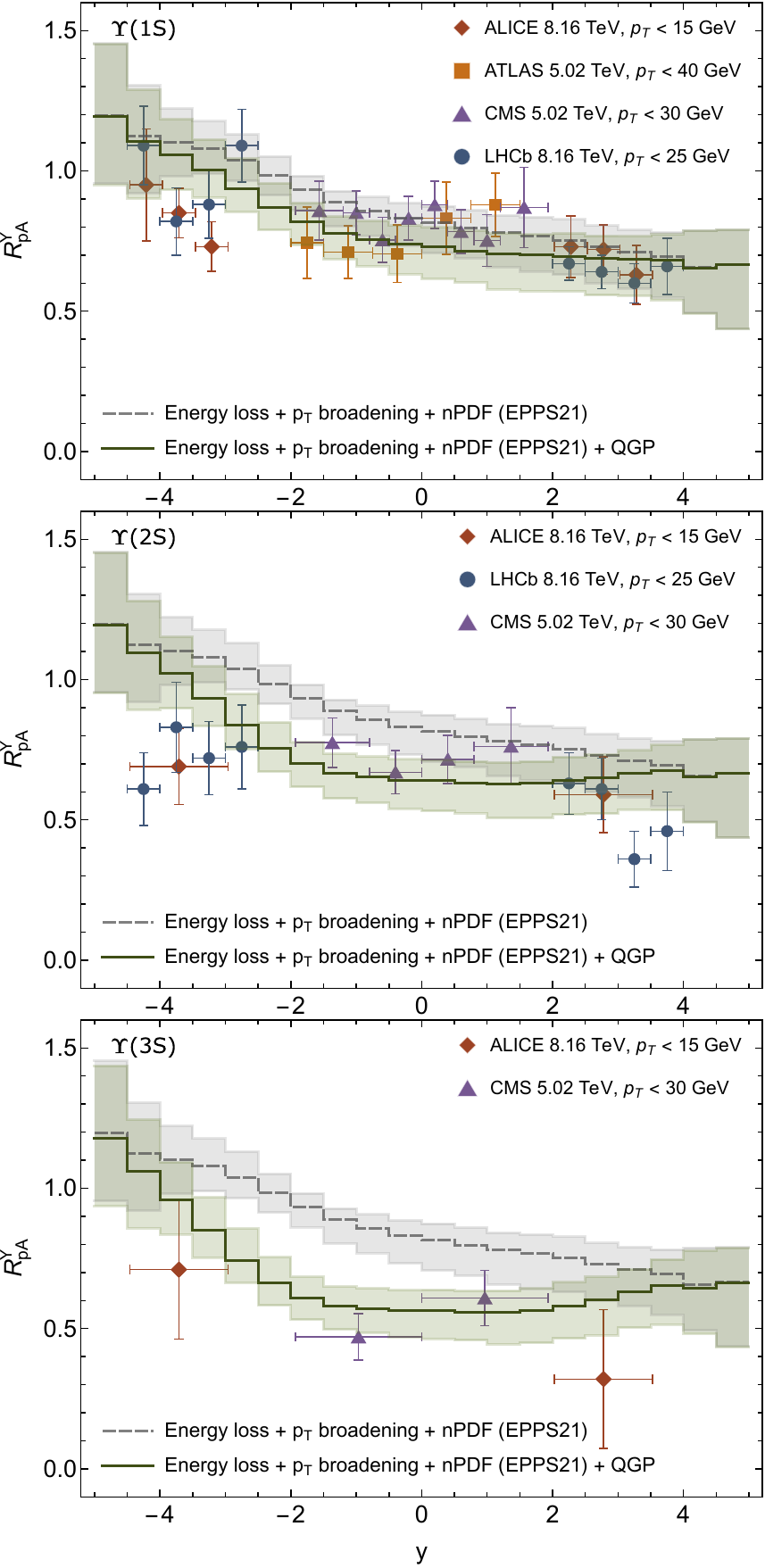}
\end{center}
\vspace{-5mm}
\caption{$\Upsilon(1S)$ (top), $\Upsilon(2S)$ (middle), and $\Upsilon(3S)$ (bottom) $R_{pA}$ as a function of $y$.  At $|y| < 2$ and $|y| \ge 2$, the model results shown are for $\sqrt{s_{NN}} = $ 5.02 TeV and \mbox{$\sqrt{s_{NN}} = $ 8.16 TeV} $p$-Pb collisions, respectively.  Horizontal error bars indicate the width of the reported rapidity bins.  The data from the ALICE, ATLAS, CMS, and LHCb Collaborations are from Refs.~\cite{ALICE:2019qie}, \cite{ATLAS:2017prf}, \cite{CMS:2022wfi}, and \cite{LHCb:2018psc}, respectively.
} 
\label{fig:rpA-y}
\end{figure}
%-----------------------------------------------

%-----------------------------------------------
\begin{figure}[t]
\begin{center}
\includegraphics[width=0.98\linewidth]{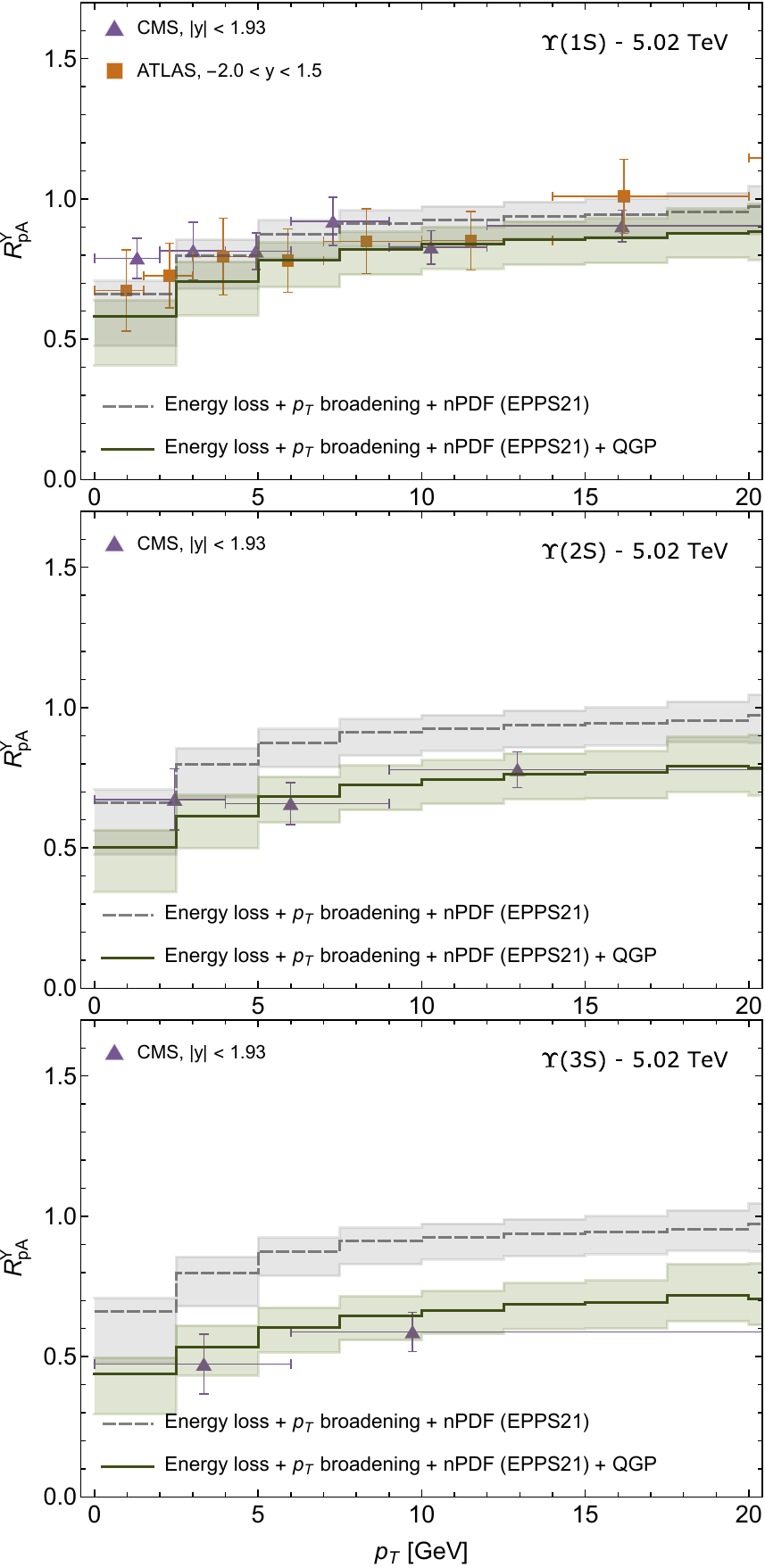}
\vspace{-5mm}
\end{center}
\caption{$\Upsilon(1S)$ (top), $\Upsilon(2S)$ (middle), and $\Upsilon(3S)$ (bottom) $R_{pA}$ as a function of $p_T$ for $\sqrt{s_{NN}} = $ 5.02 TeV for $p$-Pb collisions. Horizontal error bars indicate the width of the reported transverse momentum bins.  The CMS data points are shifted to the average momentum in each bin.  The ATLAS and CMS data are from Refs.~\cite{ATLAS:2017prf} and \cite{CMS:2022wfi}, respectively.} 
\label{fig:rpA-pt-5.02}
\end{figure}
%-----------------------------------------------

%%%%%%%%%%%%%%%%%%%%%%%%%%%%%%%%%%%%%%%%%%%%%%%%%%%%%%%%%%%%%%%%
\section{Results}
\label{sec:results}
%%%%%%%%%%%%%%%%%%%%%%%%%%%%%%%%%%%%%%%%%%%%%%%%%%%%%%%%%%%%%%%%

We now turn to our final results, which combine the effects of nPDFs, energy loss, momentum broadening, and final state QGP-induced suppression.  We present comparisons of our model predictions with data obtained at \mbox{$\sqrt{s_{NN}} = $ 5.02} and 8.16 TeV~\cite{ALICE:2019qie,ATLAS:2017prf,CMS:2022wfi,LHCb:2018psc}.  It is important to note that there are only small differences between the modifications to bottomonium production induced by all effects at these two collision energies.  In order to quantify the impact of the different collision energies, in the Appendix we present comparisons of the rapidity dependence of bottomonium suppression obtained at both collision energies, indicating the effect each of the main components in the calculation has on our results.  The results reported in this Appendix demonstrate that there are only small differences found when considering the two collision energies, however, since some subset of the data we will compare to was collected at \mbox{$\sqrt{s_{NN}} = $ 5.02 TeV} (ATLAS and CMS) and some at $\sqrt{s_{NN}} = $ 8.16 TeV (ALICE and LHCb), we have made separate computations at each collision energy in order to make appropriate comparisons.

In Fig.~\ref{fig:rpA-y} we present our results for $R^\Upsilon_{pA}$ for $\Upsilon(1S)$ (top), $\Upsilon(2S)$ (middle), and $\Upsilon(3S)$ (bottom) as a function of rapidity $y$.  In all panels, the dashed gray line is the result obtained when including nPDF effects, coherent energy loss, and momentum broadening.  The solid green line is the result obtained when including the suppression experienced by bottomonium as it traverses the QGP.  The gray shaded bands indicate the uncertainties associated with varying over the nPDF error sets.  The green-shaded bands indicate the combined effects of varying over the nPDF error sets and our assumed ranges for the heavy quarkonium transport coefficient $\hat\kappa$.  

Because the experimental collaborations report data at different collision energies, in all panels of Fig.~\ref{fig:rpA-y}, at $|y|<2$, we show our results for \mbox{$\sqrt{s_{NN}} = $ 5.02 TeV} $p$-Pb collisions, and at $|y|\geq2$, results are shown for \mbox{$\sqrt{s_{NN}} = $ 8.16 TeV} $p$-Pb collisions.  This matches the rapidity intervals of the observations of the ATLAS/CMS and ALICE/LHCb Collaborations, respectively.  As can be seen from this figure, the addition of the effect of propagation of bottomonium states through the QGP allows one to quantitatively understand the increased suppression of the $\Upsilon(2S)$ and $\Upsilon(3S)$ states, particularly at central rapidity. At the same time, the predictions for $\Upsilon(1S)$ suppression remain consistent with the experimental data.  All the states exhibit some tension with the experimental data at backward rapidity.  However, there are large experimental uncertainties in this region.  At forward rapidity, there is good agreement between our results and experimental data for $\Upsilon(1S)$ suppression; however, there is some tension with existing experimental data for the excited states.  Once again, however, we point out that there are large experimental uncertainties in this rapidity range.  The difference between our results and the experimental data could indicate the need to include additional suppression mechanisms or to make further improvements to the underlying model for QGP-induced suppression.

Turning next to the transverse momentum dependence, in Fig.~\ref{fig:rpA-pt-5.02} we present comparisons of our results for $R_{pA}$ in $\sqrt{s_{NN}} = $ 5.02 TeV $p$-Pb collisions with experimental data from the ATLAS and CMS Collaborations.  In the top, middle, and bottom panels, we show the results obtained for $\Upsilon(1S)$, $\Upsilon(2S)$, and $\Upsilon(3S)$, respectively.  The line styles and shading are the same as in Fig.~\ref{fig:rpA-y}.  We note that the theoretical calculation uses the CMS rapidity interval of $|y| < 1.93$, while the ATLAS rapidity interval for their $\Upsilon(1S)$ results shown in the top panel of Fig.~\ref{fig:rpA-pt-5.02} is slightly different.  That said, the rapidity intervals are not dramatically different, which makes such a comparison meaningful, especially considering the reported experimental and theoretical uncertainties.  As can been seen from Fig.~\ref{fig:rpA-pt-5.02}, our results are in excellent agreement with the experimental data for the transverse momentum dependence of $\Upsilon(1S)$ and $\Upsilon(2S)$ suppression.  Additionally, we see from the bottom two panels of Fig.~\ref{fig:rpA-pt-5.02} that inclusion of QGP-induced suppression is necessary to understand the experimental data.  The central values of our results for $\Upsilon(3S)$ are slightly above the experimental data.  Nonetheless, they are consistent with observations given current theoretical and experimental uncertainties.

%-----------------------------------------------
\begin{figure}[t]
\begin{center}
\includegraphics[width=0.98\linewidth]{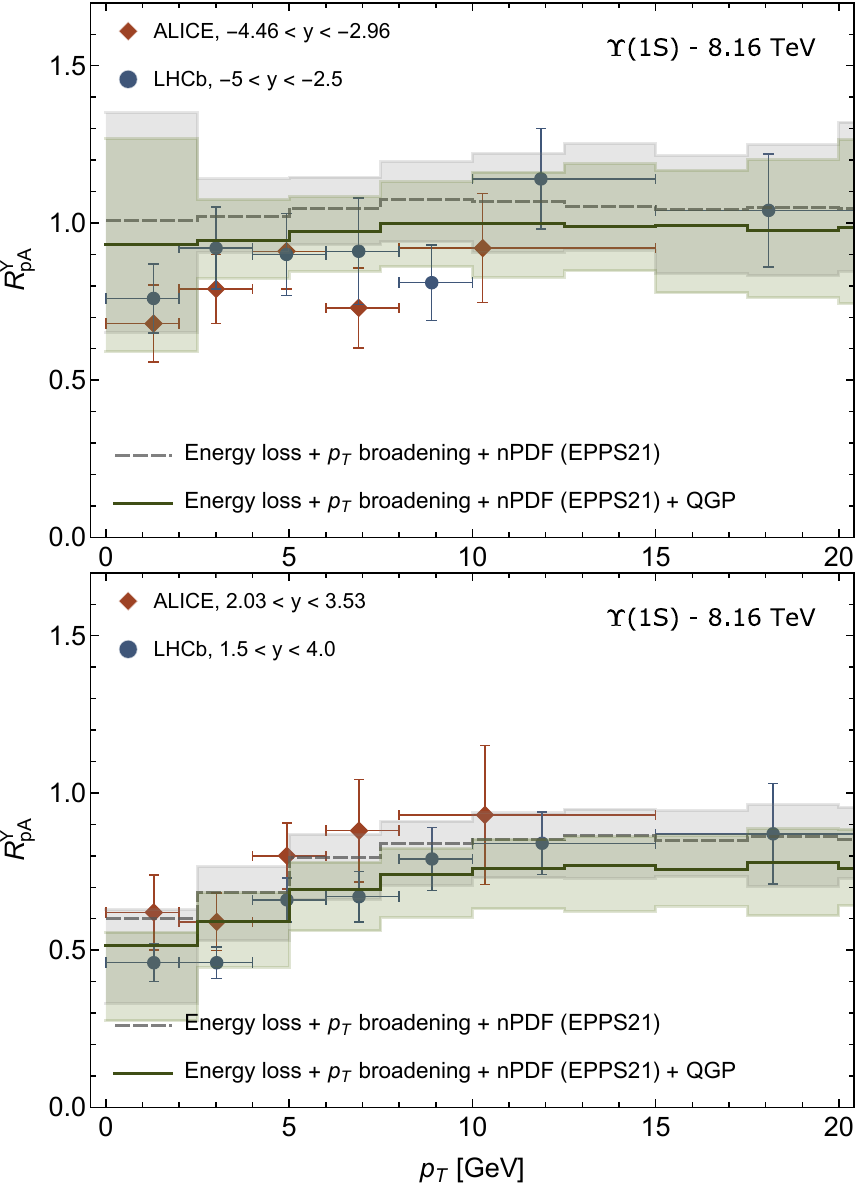}
\vspace{-4mm}
\end{center}
\caption{$\Upsilon(2S)$ $R_{pA}$ as a function of $p_T$ for \mbox{$\sqrt{s_{NN}} = $ 8.16 TeV}.  Horizontal error bars indicate the reported transverse momentum bins.  All data points are shifted to the average momentum in each bin.  The LHCb Collaboration data are from Ref.~\cite{LHCb:2018psc}.} 
\caption{$\Upsilon(1S)$ $R_{pA}$ as a function of $p_T$ for \mbox{$\sqrt{s_{NN}} = $ 8.16 TeV}.  Horizontal error bars indicate the width of the reported transverse momentum bins.  All data points are shifted to the average momentum in each bin. The ALICE and LHCb data are from Refs.~\cite{ALICE:2019qie} and \cite{LHCb:2018psc}, respectively.} 
\label{fig:rpA-1S-pt-8.16}
\end{figure}
%-----------------------------------------------

%-----------------------------------------------
\begin{figure}[t]
\begin{center}
\includegraphics[width=0.98\linewidth]{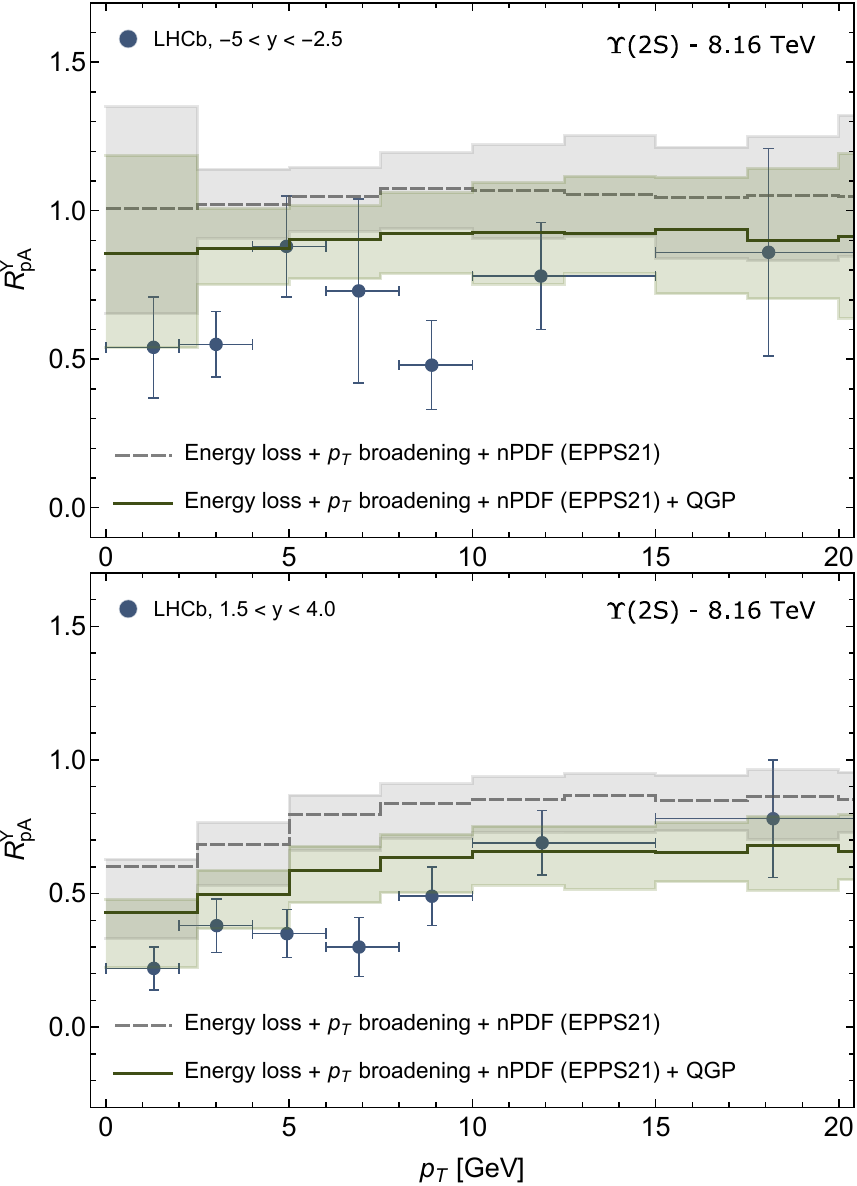}
\vspace{-4mm}
\end{center}
\caption{$\Upsilon(2S)$ $R_{pA}$ as a function of $p_T$ for \mbox{$\sqrt{s_{NN}} = $ 8.16 TeV}.  Horizontal error bars indicate the reported transverse momentum bins.  All data points are shifted to the average momentum in each bin.  The LHCb Collaboration data are from Ref.~\cite{LHCb:2018psc}.} 
\label{fig:rpA-2S-pt-8.16}
\end{figure}
%-----------------------------------------------

Next, in Fig.~\ref{fig:rpA-1S-pt-8.16} we present comparisons of our results for $\Upsilon(1S)$ $R_{pA}$ in $\sqrt{s_{NN}} = $ 8.16 TeV $p$-Pb collisions with experimental data from the ALICE and LHCb Collaborations.  The line styles and shading are the same as in the previous two figures.  In the top and bottom panels, we show results obtained at backward and forward rapidities, respectively.  The theoretical calculations at backward rapidity are in the interval $-5 < y < -2.5$, and those at forward rapidity are in the range $1.5 < y < 4$.  Both cases correspond to the intervals used by the LHCb Collaboration.  Once again, although the rapidity intervals used by the ALICE Collaboration are slightly different, this comparison is still meaningful.  As can be seen from the top panel of Fig.~\ref{fig:rpA-1S-pt-8.16}, inclusion of QGP-induced suppression brings our results into better agreement with experimental data collected at backward rapidity; however, there are indications that in this kinematic region there may be additional suppression not accounted for by our calculations.  In the bottom panel, we find quite reasonable agreement between our results and the experimental data for the transverse momentum dependence of $\Upsilon(1S)$ suppression.

Finally, in Fig.~\ref{fig:rpA-2S-pt-8.16} we present comparisons of our results for $\Upsilon(2S)$ $R_{pA}$ in $\sqrt{s_{NN}} = $ 8.16 TeV $p$-Pb collisions with experimental data from the LHCb Collaboration.  The line styles and shading are the same as in the previous three figures.  As before, in the top and bottom panels, we show results obtained at backward and forward rapidities, respectively.  The top panel shows that inclusion of final-state effects brings the results into better agreement with LHCb data; however, there are indications that for backward rapidity, we slightly underestimate the observed suppression.  The bottom panel shows that inclusion of final-state effects results in better agreement with experimental data at forward rapidity; however, our results are slightly above the experimental observations at low transverse momentum.  This is to be contrasted with the results shown in Fig.~\ref{fig:rpA-pt-5.02} for central rapidities which seem to explain the CMS $\Upsilon(2S)$ suppression data very well.  As stated above, this implies that there could be additional suppression mechanisms, such as hadronic comovers or corrections beyond NLO in the OQS+pNRQCD treatment that are important in the far-backward or far-forward rapidity regions.

%%%%%%%%%%%%%%%%%%%%%%%%%%%%%%%%%%%%%%%%%%%%%%%%%%%%%%%%%%%%%%%%
\section{Conclusions and outlook}
\label{sec:conclusions}
%%%%%%%%%%%%%%%%%%%%%%%%%%%%%%%%%%%%%%%%%%%%%%%%%%%%%%%%%%%%%%%%

In this paper we considered the suppression of bottomonium in min-bias $\sqrt{s_{NN}} = 5.02$ TeV and 8.16 TeV $p$-Pb collisions, taking into account nPDF effects, coherent energy loss, momentum broadening, and final-state interactions of bottomonium with the QGP.  We found that incorporation of all of these effects provides a reasonably accurate description of available experimental data, given the current experimental and theoretical uncertainties. Specifically, for $\Upsilon(1S)$, final-state interactions with the QGP represent a small correction to the effects of nPDFs, coherent energy loss, and transverse momentum broadening. However, for $\Upsilon(2S)$ and $\Upsilon(3S)$, including QGP-induced suppression is essential to achieve agreement with the available data. This further supports the idea that a hot, short-lived, QGP is created in min-bias \mbox{$p$-Pb} collisions. This conclusion is consistent with previous studies of charmonium suppression \cite{Du:2018wsj,Wen:2022utn,Chen:2023toz}, where including QGP-induced suppression was necessary to understand the increased suppression of $\psi(2S)$ compared to $J/\psi$.

Looking to the future, there are many avenues for improving the work presented in this paper.  First, one may consider varying the different model components, including using different nPDF sets, considering different models of the QGP-induced suppression such as transport models including regeneration or phenomenological complex-potential models, including a more realistic underlying potential model in the OQS+pNRQCD calculation of quarkonium suppression \cite{magorsch}, and/or taking into account the effects of comoving hadrons in the kinematic and spatiotemporal regions where it is called for.  Additionally, the centrality dependence of a limited number of bottomonium observables is available in $p$-Pb collisions, so one may also consider the centrality dependence in these cases.  Another avenue for future research includes the application of our methodology to the suppression of charmonium.  In this case, regeneration will need to be considered in order to quantitatively assess its impact on QGP-induced suppression.  Finally, we note that there exists a straightforward means of applying the methods employed here to $AA$ collisions in order to understand the interplay of cold and hot nuclear matter effects in such collisions and draw firmer conclusions concerning the phenomenological extraction of heavy quarkonium transport coefficients in hot QCD from experimental observations.

\acknowledgments{
We thank R. Rapp for encouragement.  This work was supported by the U.S. Department of Energy, Office of Science, Office of Nuclear Physics through the Topical Collaboration in Nuclear Theory on Heavy-Flavor Theory (HEFTY) for QCD Matter under contract number DE-SC0023547.  M.S.\ and S.T.\ were also supported by the U.S.\ Department of Energy, Office of Science, Office of Nuclear Physics (Nuclear Theory) under contract number DE-SC0013470.  R.V.\ was also supported by the US Department of Energy by LLNL under contract DE-AC52-07NA27344 and by the US Department of Energy, Office of Science, Office of Nuclear Physics (Nuclear Theory) under contract number DE-SC-0004014.
}

\appendix

%-----------------------------------------------
\begin{figure}[b]
\begin{center}
\includegraphics[width=0.95\linewidth]{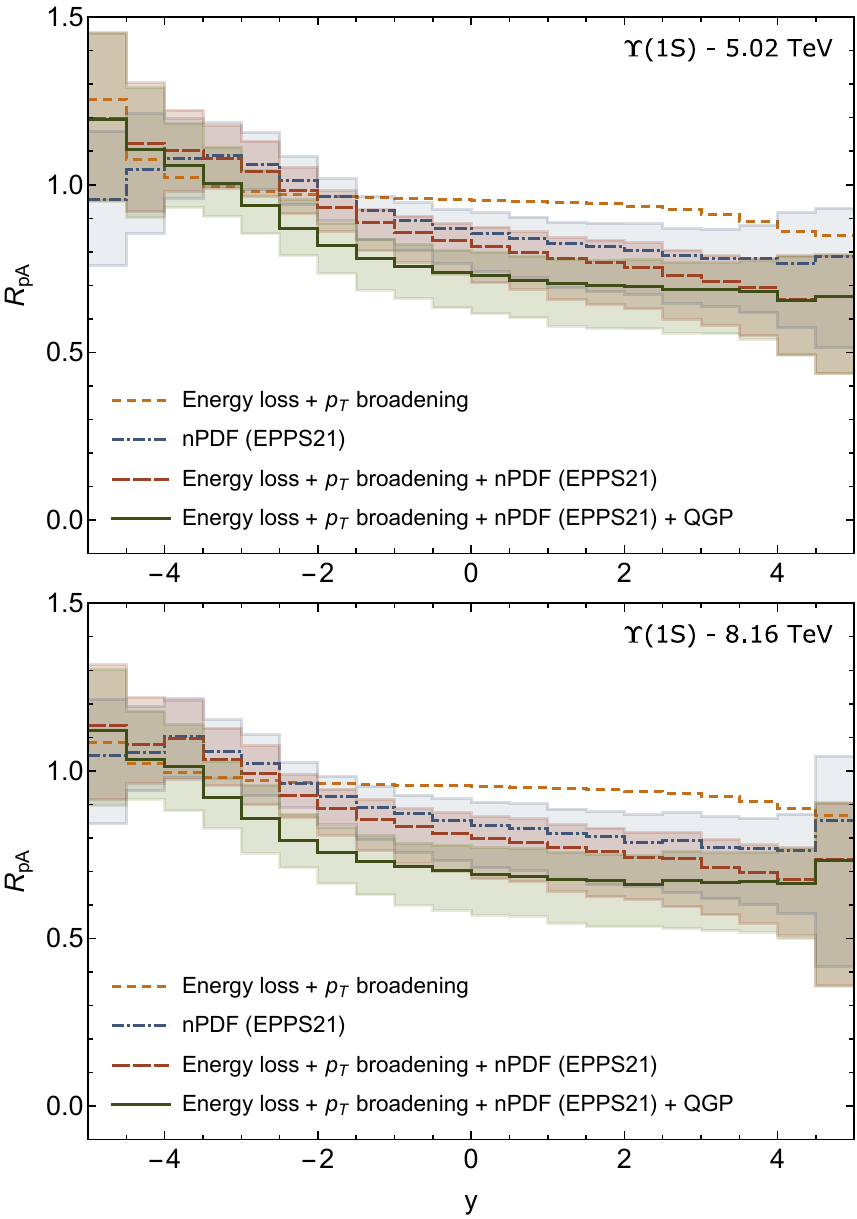}
\end{center}
\vspace{-5mm}
\caption{Different effects contributing to $\Upsilon(1S)$ $R_{pA}$ as a function of $y$ for $\sqrt{s_{NN}} = $ 5.02 TeV (top) and $\sqrt{s_{NN}} = $ 8.16 TeV (bottom).} 
\label{fig:rpA-Y1S-y-5v8}
\end{figure}
%-----------------------------------------------

%-----------------------------------------------
\begin{figure*}[t]
\begin{center}
\includegraphics[width=0.48\linewidth]{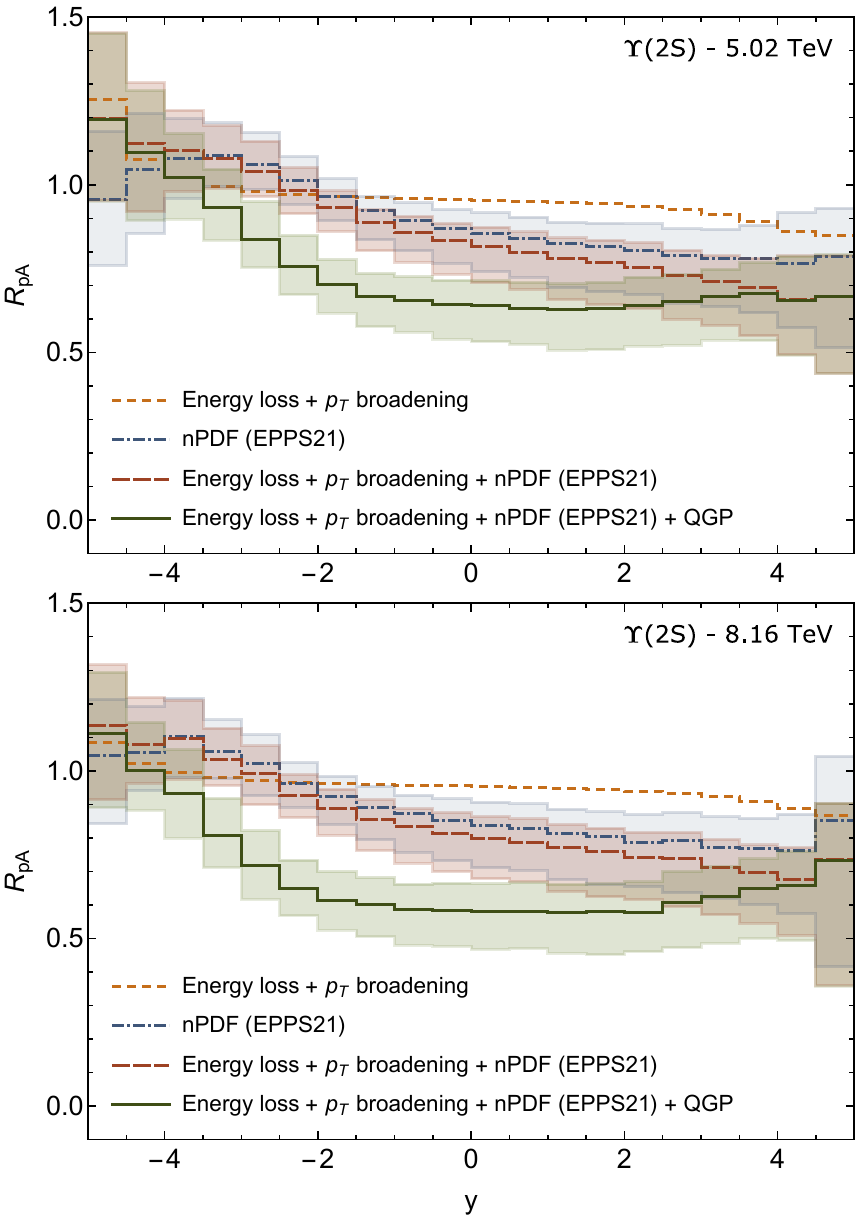}
\;\;
\includegraphics[width=0.48\linewidth]{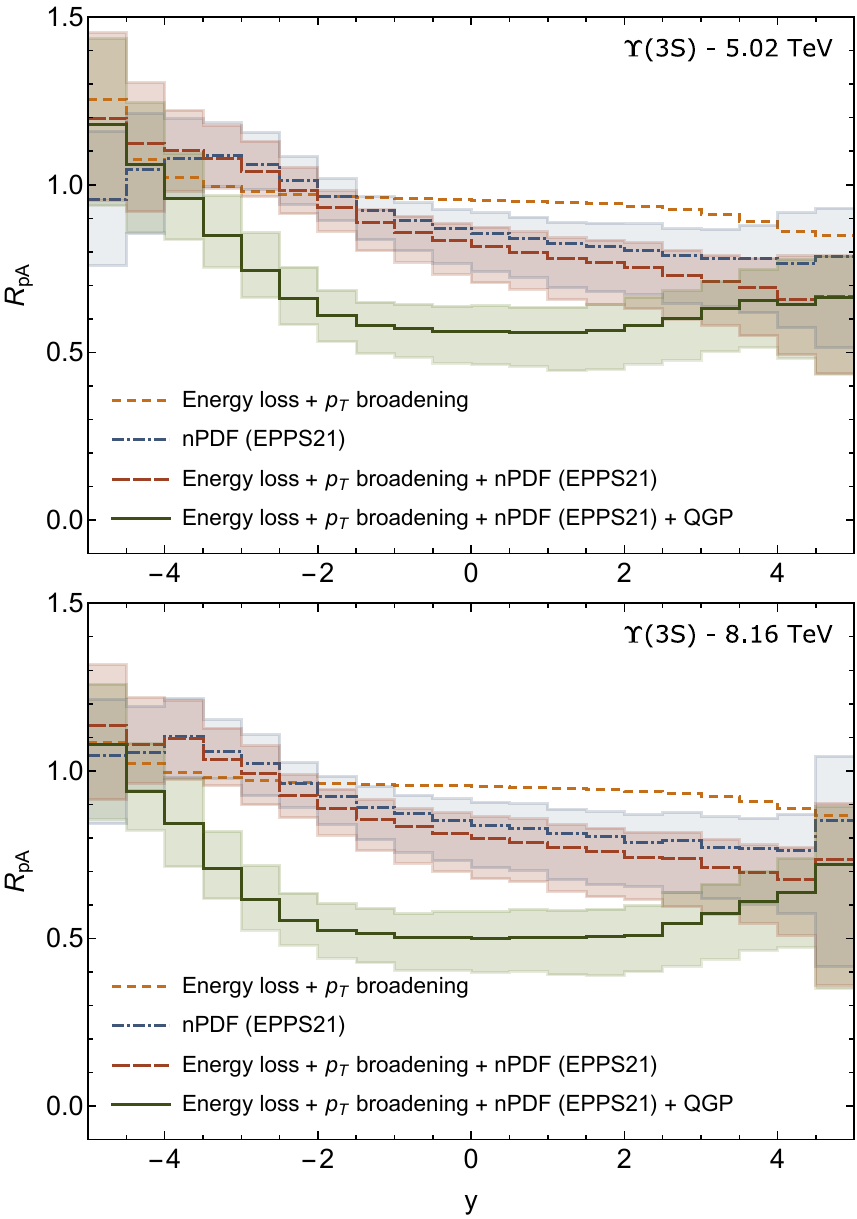}
\end{center}
\vspace{-5mm}
\caption{Different effects contributing to $\Upsilon(2S)$  (left) and $\Upsilon(3S)$ (right) $R_{pA}$ as a function of $y$ for $\sqrt{s_{NN}} = $ 5.02 TeV (top) and $\sqrt{s_{NN}} = $ 8.16 TeV (bottom).} 
\label{fig:rpA-Y2S-y-5v8}
\end{figure*}
%-----------------------------------------------

%%%%%%%%%%%%%%%%%%%%%%%%%%%%%%%%%%%%%%%%%%%%%%%%%%%%%%%%%%%%%%%%
\section*{APPENDIX: COMPARISONS OF RESULTS AT 5.02 TEV AND 8.16 TEV}
\label{app:58comp}
%%%%%%%%%%%%%%%%%%%%%%%%%%%%%%%%%%%%%%%%%%%%%%%%%%%%%%%%%%%%%%%%

In this Appendix we present comparisons of all effects considered on the $\Upsilon(1S)$, $\Upsilon(2S)$, and $\Upsilon(3S)$ at \mbox{$\sqrt{s_{NN}} = $ 5.02 TeV} and $\sqrt{s_{NN}} = $ 8.16 TeV.  While these results are combined in Fig.~\ref{fig:rpA-y}, here they are shown separately so that the reader can see the effect of varying the collision energy on each effect considered.  In Fig.~\ref{fig:rpA-Y1S-y-5v8} we show our results for the $\Upsilon(1S)$ $R_{pA}$ resulting from energy loss and $p_T$ broadening (dashed orange line), nPDF effects (dot dashed blue line), the combination of energy loss plus momentum broadening plus nPDF effects (dashed red line), and finally all effects combined including QGP-induced suppression (solid green line).  The top panel contains the results obtained at \mbox{$\sqrt{s_{NN}} = $ 5.02 TeV} and the bottom panel contains the results obtained at $\sqrt{s_{NN}} = $ 8.16 TeV.  The shaded bands indicate the theoretical uncertainties associated with varying over the nPDF error sets on the results with the nPDFs alone and those including energy loss and momentum broadening.  In the complete result (solid green line), the bands indicate the combined uncertainty resulting from varying over both the nPDF error sets and the assumed range of the QTraj transport coefficient $\hat\kappa$ (see Fig.~\ref{fig:rpA-y-qtraj-5.02} for typical uncertainties associated with this variation).

Figure~\ref{fig:rpA-Y2S-y-5v8} shows the same for the $\Upsilon(2S)$ (left) and $\Upsilon(3S)$ (right) states.  As can be seen in Figs.~\ref{fig:rpA-Y1S-y-5v8} and \ref{fig:rpA-Y2S-y-5v8}, there are only small differences in the suppression computed at each collision energy.  In the main body of the text we have merged the results at the different collision energies in Fig.~\ref{fig:rpA-y}, with $|y|<2$ corresponding to our results obtained at $\sqrt{s_{NN}} = $ 5.02 TeV and $|y| \geq 2$ corresponding to our results obtained at \mbox{$\sqrt{s_{NN}} = $ 8.16 TeV}.

\bibliographystyle{apsrev4-1}
\bibliography{main-inspire,main-notinspire}

\end{document}